\documentclass[aps,prd,groupedaddress,preprintnumbers,
              floatfix, nofootinbib,longbibliography]{revtex4}
\usepackage{textcomp}
\usepackage[latin1]{inputenc}                    
\usepackage{graphicx}                            
\usepackage{epsfig}
\usepackage[export]{adjustbox}
\usepackage{mathrsfs}
\usepackage{mathtools}

\usepackage{epsf}
\usepackage{latexsym}                            
\usepackage{amsfonts}                            
\usepackage{amssymb}                             
\usepackage{amsmath}                             
\usepackage{slashed}
\usepackage[mathscr]{eucal}                      
\usepackage{dcolumn}                             
\usepackage{multirow}                               
\usepackage{bm}                                  
\usepackage{hyperref}                            
\usepackage[usenames]{color}
\usepackage[dvipsnames]{xcolor}

\usepackage{color}
\usepackage{ulem}

\graphicspath{{.}{Graphics/}}                    
\newcommand{\msbar}{\overline{\mbox{\rm MS}}}  
\newcommand{\mmsbar}{{\rm M}\overline{\mbox{\rm MS}}}  

\newcommand{\MSb}{\overline{\mathrm{MS}}}

\newcommand{\be}{\begin{equation}}
\newcommand{\ee}{\end{equation}}
\newcommand{\bea}{\begin{eqnarray}}
\newcommand{\eea}{\end{eqnarray}}

\def \3{\ss }

\newcommand{\beqn}{\begin{eqnarray}}
\newcommand{\eeqn}{\end{eqnarray}}

\newcommand{\UV}{\text{\tiny{UV}}}
\newcommand{\IR}{\text{\tiny{IR}}}

\def\poz{b}
\def\teml{a}
\def\bon{c}

\begin{document}

\begin{titlepage}
  {\vspace{-0.5cm} \normalsize
  \hfill \parbox{60mm}{
}}\\[10mm]
  \begin{center}
    \begin{LARGE}
            \textbf{The role of zero-mode contributions \\[0.2cm]
            in the matching for the twist-3 PDFs $e(x)$ and $h_{L}(x)$ \\[0.2cm] }
    \end{LARGE}
  \end{center}

\vspace*{1cm}

 \vspace{-0.8cm}
  \baselineskip 20pt plus 2pt minus 2pt
  \begin{center}
    \textbf{
      Shohini Bhattacharya$^{(\teml)}$,
      Krzysztof Cichy$^{(\poz)}$,
      Martha Constantinou$^{(\teml)}$,\\
      Andreas Metz$^{(\teml)}$,
      Aurora Scapellato$^{(\poz)}$,
      Fernanda Steffens$^{(\bon)}$
      }
\end{center}

  \begin{center}
    \begin{footnotesize}
      \noindent 	
    $^{(\teml)}$ {\it Temple University, 1925 N.~12th Street, Philadelphia, PA 19122-1801, USA} \\
 	$^{(\poz)}$  {\it Faculty of Physics, Adam Mickiewicz University, Uniwersytetu Pozna\'{n}skiego 2, 61-614 Pozna\'{n}, Poland} \\
 	$^{(\bon)}$ {\it Institut f\"{u}r Strahlen- und Kernphysik, Rheinische Friedrich-Wilhelms-Universit\"{a}t Bonn,  \\ 
 	 Nussallee 14-16, 53115 Bonn, Germany} \\
     \vspace{0.2cm}
    \end{footnotesize}
  \end{center}

\centerline{\today}

\begin{abstract}
The perturbative procedure of matching was proposed to connect parton quasi-distributions that are calculable in lattice QCD to the corresponding light-cone distributions which enter physical processes.
Such a matching procedure has so far been limited to the twist-2 distributions. Recently, we addressed the matching for the twist-3 PDF $g_T(x)$. 
In this work, we extend our perturbative calculations to the remaining twist-3 PDFs, $e(x)$ and $h_{L}(x)$. 
In particular, we discuss the non-trivialities involved in the calculation of the singular zero-mode contributions for the quasi-PDFs. 
\end{abstract}

\maketitle
\end{titlepage}

\section{Introduction}
\label{s:introduction}

The twist-3 parton distribution functions (PDFs) $e (x)$ and $h_L (x)$ were introduced some 30 years ago~\cite{Jaffe:1991kp, Jaffe:1991ra}.
They complement the twist-3 PDF $g_T(x)$, which enters the cross section of polarized deep-inelastic lepton-nucleon scattering (DIS).
Twist-3 PDFs are of general interest as they contain information about quark-gluon-quark correlations in the nucleon~\cite{Balitsky:1987bk, Kanazawa:2015ajw}.
Moreover, a semi-classical relation between the function $e(x)$ and the (transverse) force acting on transversely polarized quarks in an unpolarized nucleon has been reported in Ref.~\cite{Burkardt:2008ps}.
In Ref.~\cite{Seng:2018wwp}, $e(x)$ was shown to be related to the poorly-known hadronic matrix element of the quark chromo-magnetic dipole moment operator, which is an essential input in the study of nuclear electric dipole moments (EDMs), and hence this connection can provide new insight into physics beyond the standard model.
Recently, the role of $e(x)$ has also been discussed in relation to the mass structure of hadrons~\cite{Hatta:2020iin} (see, also Ref.~\cite{Ji:2020baz}).
Unlike $g_T(x)$, both $e(x)$ and $h_L(x)$ are chiral odd and hence can only show up in observables with other chiral-odd functions. 
This feature makes it challenging to extract information on these functions from experiment. 
In Ref.~\cite{Jaffe:1991ra}, it was argued that $e(x)$ can be accessed in an unpolarized Drell-Yan process, but only at the level of twist-4. Soon after, it was shown that $e(x)$ could also be measured through a particular twist-3 single-spin asymmetry in semi-inclusive DIS~\cite{Levelt:1994np}, which has been measured by the HERMES and CLAS collaborations~\cite{Airapetian:1999tv, Airapetian:2001eg, Avakian:2003pk, Gohn:2014zbz}.  
An alternative process for addressing $e(x)$ is di-hadron production in electron-proton collisions~\cite{Bacchetta:2003vn}.
A first attempt to extract information about $e(x)$ through this channel, based on preliminary data from the CLAS collaboration, can be found in Ref.~\cite{Courtoy:2014ixa}.
A twist-3 double-spin asymmetry in the Drell-Yan process could be used to address $h_L(x)$~\cite{Jaffe:1991kp, Jaffe:1991ra, Tangerman:1994bb, Koike:2008du}, and other final states in polarized hadronic collisions could in principle be considered as well --- see, for instance, the discussion in Refs.~\cite{Liang:2012rb, Metz:2012fq}.
But so far no information exists on $h_{L}(x)$ from the experimental side.

\subsection{Delta function singularities in $e(x)$ and $h_L(x)$}
An interesting and sometimes controversially-discussed feature of $e(x)$ and $h_L(x)$ regards the possible existence of singular zero-mode ($x = 0$) contributions, that is, delta-function singularities ($\delta(x)$), and their implication on sum rules. 
For the sake of this discussion, we summarize below the sum rules for the lowest moments of $e(x)$ and $h_L(x)$.
By definition, the lowest moment of the the flavor-singlet combination of $e(x)$ gives the pion-nucleon sigma term $\sigma_{\pi N}$~\cite{Jaffe:1991ra}, 
\begin{equation}
\int^{1}_{-1} dx (e^{u} (x) + e^{d} (x)) = \dfrac{\sigma_{\pi N}}{m} \, ,
\label{e:sigma_sum_rule}
\end{equation}
where,
\begin{equation}
\sigma_{\pi N} =  \frac{m}{2M_{N}}  \langle P | \, \big ( \bar{\psi}^{u} (0) \psi^{u} (0) +  \bar{\psi}^{d} (0) \psi^{d} (0) \big ) \, | P \rangle  \, , \quad  m= \dfrac{1}{2} (m_{u} + m_{d})\, ,
\label{e:sigma_sum_rule}
\end{equation}
and $M_{N}$ is the nucleon mass. On the basis of rotational invariance, it was shown that the lowest moments of $h_L(x)$ and the twist-2 transversity $h_1(x)$~\cite{Burkardt:1995ts, Tangerman:1994bb} are connected as
\begin{equation}
\int^{1}_{-1} dx \,h_{L} (x) = \int^{1}_{-1} dx\, h_{1} (x)\,,
\label{e:h sum rule}
\end{equation}
which is the counterpart of the Burkhardt-Cottingham sum rule that relates $g_T(x)$ and the (twist-2) helicity distribution $g_1(x)$~\cite{Burkhardt:1970ti}.

As mentioned above, there has been discussion on whether one can get around the presence of the zero modes. 
Refs.~\cite{Kodaira:1998jn, Efremov:2002qh, Pasquini:2018oyz} emphasized that a $\delta(x)$ singularity in $e(x)$ is a consequence of the QCD equation of motion (EOM). Specifically, one can split $e(x)$ as
\begin{equation}
e^{q}(x) = \dfrac{\delta(x)}{2M_{N}} \langle P | \bar{\psi}^{q} (0) \psi^{q} (0) | P \rangle + \tilde{e}^{q}(x) + e^{q}_{m}(x) \,,
\label{eq:e_decomp}
\end{equation} 
where $\tilde{e}$ is a ``pure" twist-3 term (which encodes quark-gluon-quark interactions) and $e_{m}$ is a current-mass term.
Using the decomposition of Eq.~(\ref{eq:e_decomp}) in the above mentioned sum rule, one finds
\begin{equation}
\int dx \,\tilde{e}^{q}(x) = 0 \,,  \qquad \int dx \,\tilde{e}_{m}(x) = 0 \,,
\label{e:e_tilde_mass_sum_rules}
\end{equation}
which implies that the first moment of $e(x)$ entirely receives contribution from the $\delta(x)$ term. Very recently, it was argued, again on the grounds of EOM approach, that the coefficient of $\delta(x)$ is zero~\cite{Ma:2020kjz}. A critique on that work was drawn in Ref.~\cite{Hatta:2020iin}, ruling out the possibility of a cancellation of $\delta(x)$ in $e(x)$. By reconstructing $h_L(x)$ from its operator product expanded (OPE) form, Ref.~\cite{Jaffe:1991ra} showed that $h_L(x)$ comprised of three terms: a twist-2 term, a ``pure" twist-3 term, and a current-mass term. Through a foreseeable discontinuity in the integral relation between $h_L(x)$ and the mass term, Ref.~\cite{Burkardt:2001iy} indicated the existence of a possible $\delta(x)$ in $h_L(x)$. The need for such a singularity was also justified for a compliance with the sum rule mentioned in Eq.~(\ref{e:h sum rule}) as the twist-2 part, $h_1(x)$, is continuous at $x=0$.

The first attempt to calculate $e(x)$ and $h_L(x)$ was made in the MIT bag model~\cite{Jaffe:1991ra,Signal:1996ct}. However, no $\delta(x)$ singularity was found. Calculations in diquark spectator models, with form factors, did also not indicate such singularities~\cite{Jakob:1997wg}. A recent study in the same (spectator) model~\cite{Aslan:2018tff}, using a cut-off for the transverse momentum integration instead of form factors, showed that a $\delta(x)$ is present in both $e(x)$ and $h_L(x)$. 
A $\delta(x)$ contribution in $e(x)$ was also found in non-perturbative calculations in the chiral quark-solition Model ($\chi$QSM)~\cite{Wakamatsu:2003uu,Schweitzer:2003uy,Ohnishi:2003mf,Cebulla:2007ej}. Remarkably, the coefficient of $\delta(x)$ was shown to be related to the sigma term and therefore the vacuum quark condensate -- a quantity directly related to the chiral symmetry breaking in the QCD vacuum. 
Thus, this important finding demonstrated that the non-trivial structure of the QCD vacuum, encoded in the condensate, can show up in a physical observable. To shed some light into the mechanism responsible for the singularities in the twist-3 PDFs, an interesting calculation in the (1+1)-dimensional Gross-Neveu model was presented in Ref~\cite{Burkardt:1995ts}. 
The origin of $\delta(x)$ was identified to be due to the long range quark-quark correlations, which in fact is the same mechanism responsible for $\delta(x)$ in $e(x)$ in $\chi$QSM. 
One-loop perturbative calculations of $e(x)$ and $h_L(x)$ in quark target models~\cite{Burkardt:2001iy,Aslan:2018tff} also indicated the presence of $\delta(x)$. Interestingly, in calculations employing the light-front Hamiltonian approach instead of the Feynman-diagram approach, as in Refs.~\cite{Burkardt:2001iy,Aslan:2018tff}, no such singularities were observed in $e(x)$~\cite{Mukherjee:2009uy} and $h_L(x)$~\cite{Kundu:2001pk}, which can well be due to an insufficiency of the used approach to deal with zero modes.
Generally, it is accepted that sum rules like in Eq.~(\ref{e:h sum rule}) are violated if $\delta(x)$ contributions are not included in the twist-3 PDFs~\cite{Kundu:2001pk, Burkardt:2001iy, Aslan:2018tff}. 
We note in passing that zero-mode contributions can also generate discontinuities for higher-twist generalized parton distributions~\cite{Aslan:2018zzk, Aslan:2018tff}, thus endangering factorization of certain observables in hard exclusive reactions.

\subsection{Accessing PDFs from lattice QCD}
\label{s:introduction_quasi}
By now, we already see that there are various theoretical statements available in the literature about the $\delta(x)$ singularities, with some of them being contradictory.
Lattice QCD calculations with appropriate lattice parameters close to the continuum limit and with large volumes, may be able to offer some insights on the above matter in the future. However, the explicit time-dependence of the light-cone PDFs  prohibits their direct calculation on Euclidean lattices.
In 2013, there was a breakthrough proposal by Ji to calculate instead parton quasi-distributions (quasi-PDFs)~\cite{Ji:2013dva,Ji:2014gla}. Quasi-PDFs are defined in terms of spatial correlation functions of fast-moving hadrons, and therefore can be directly calculated on Euclidean lattices. At large, but finite, momentum, such correlation functions can be matched to their respective light-cone PDFs prior to the UV renormalization. On the lattice, one is constrained to apply the UV renormalization before taking the infinite momentum limit. The issue of the limits leads to differences in the UV behavior between the light-cone PDFs and the quasi-PDFs. 
The key underlying idea of this approach is that the non-perturbative physics should be the same for the light-cone and the quasi-PDFs.
The differences in the UV behavior can be calculated and rectified perturbatively in Large Momentum Effective Theory (LaMET), through a procedure known as matching~\cite{Xiong:2013bka,Stewart:2017tvs,Izubuchi:2018srq}. 
Apart from the quasi-PDF approach as a way to directly access the $x$-dependence of the PDFs in lattice QCD, several other ideas have been put forth~\cite{Braun:1994jq,Detmold:2005gg,Braun:2007wv,Ma:2014jla,Chambers:2017dov,Hansen:2017mnd,Radyushkin:2017cyf,Orginos:2017kos,Ma:2017pxb,Radyushkin:2017lvu,Liang:2017mye,Detmold:2018kwu}.

In the last few years, there has been significant advances, both in theory and in lattice QCD. This includes the proof of renormalizability~\cite{Ji:2015jwa,Ishikawa:2016znu,Chen:2016fxx}, the development of a renormalization prescription~\cite{Constantinou:2017sej, Alexandrou:2017huk}, which was extensively implemented on the lattice~\cite{Chen:2017mzz, Ji:2017oey, Ishikawa:2017faj, Green:2017xeu, Spanoudes:2018zya, Zhang:2018diq, Li:2018tpe, Constantinou:2019vyb}. A plethora of other aspects regarding quasi-PDFs and Euclidean correlators in general have also been extensively studied~\cite{Ji:2014hxa, Li:2016amo, Monahan:2016bvm, Radyushkin:2016hsy, Radyushkin:2017ffo, Carlson:2017gpk, Briceno:2017cpo, Xiong:2017jtn, Rossi:2017muf, Ji:2017rah, Wang:2017qyg, Chen:2017mie, Monahan:2017hpu, Radyushkin:2018cvn, Zhang:2018ggy, Ji:2018hvs, Xu:2018mpf, Jia:2018qee, Briceno:2018lfj, Rossi:2018zkn, Radyushkin:2018nbf, Ji:2018waw, Karpie:2018zaz, Braun:2018brg, Liu:2018tox, Ebert:2018gzl, Briceno:2018qfa, Ebert:2019okf}. The first lattice results for quasi-PDFs and other related quantities constitute an important development in this field~\cite{Lin:2014zya, Alexandrou:2015rja, Chen:2016utp, Alexandrou:2016jqi, Zhang:2017bzy, Alexandrou:2017huk, Chen:2017mzz, Green:2017xeu, Lin:2017ani, Orginos:2017kos, Bali:2017gfr, Alexandrou:2017dzj, Chen:2017gck, Alexandrou:2018pbm, Chen:2018fwa, Alexandrou:2018eet, Liu:2018uuj, Bali:2018spj, Lin:2018qky, Fan:2018dxu, Bali:2018qat, Sufian:2019bol, Alexandrou:2019lfo,Izubuchi:2019lyk,Cichy:2019ebf,Joo:2019jct,Joo:2019bzr,Alexandrou:2019dax,Chai:2020nxw,Joo:2020spy,Bhat:2020ktg}. 
Additionally, the verification of convergence of quasi-PDFs to their light-cone counterparts in model calculations further substantiate these quasi-distributions to be reliable tools to study the light-cone PDFs~\cite{Gamberg:2014zwa,Bacchetta:2016zjm,Nam:2017gzm,Broniowski:2017wbr,Hobbs:2017xtq, Broniowski:2017gfp,Xu:2018eii,Bhattacharya:2018zxi,Bhattacharya:2019cme,Son:2019ghf,Ma:2019agv,Kock:2020frx,Luo:2020yqj}. We refer to~\cite{Monahan:2018euv,Cichy:2018mum,Zhao:2018fyu,Ji:2020ect,Constantinou:2020pek} for an up-to-date compendium of progress in the field of studying light-cone PDFs through Euclidean correlators in lattice QCD. 

The procedure of matching has largely been explored for the twist-2 distribution functions~\cite{Xiong:2013bka,Ji:2015jwa,Ji:2015qla,Xiong:2015nua,Wang:2017qyg,Stewart:2017tvs,Radyushkin:2018cvn,Zhang:2018ggy,Izubuchi:2018srq,Liu:2018tox,Liu:2019urm,Wang:2019tgg,Wang:2019msf,Radyushkin:2019owq,Balitsky:2019krf}. Recently, we computed the first ever one-loop matching equations for the twist-3 PDF $g_T (x)$~\cite{Bhattacharya:2020xlt}, which we implemented on lattice data in Ref.~\cite{Bhattacharya:2020cen}. Here, we extend our work, for the case of $e(x)$ and $h_L(x)$. Specifically, we calculate  the light-cone PDFs $e(x)$ and $h_L(x)$, and their quasi-PDFs conterparts, $e_{\rm Q}$ and $h_{L, \rm Q}$, in a quark target to one-loop order in perturbative QCD (pQCD). 
The ultimate goal of this work is to obtain the appropriate matching equations.
We anticipate that a full matching formula will also involve mixing with quark-gluon-quark correlators. 
In the present work, we do not consider such mixing.

\vspace*{0.2cm}
\centerline{------------------}
\vspace*{0.2cm}

We organize the manuscript as follows: In Sec.~\ref{s:definitions} we provide the definition of the light-cone PDFs $e(x)$ and $h_{L}(x)$, and of the corresponding quasi-PDFs $e_{\rm Q}(x)$ and $h_{L, \rm Q}(x)$. In Sec.~\ref{s:one_loop_results}, we present one-loop pQCD results for $e(x)$ ($e_{\rm Q}(x)$) and $h_{L}(x)$ ($h_{L, \rm Q}(x)$) in the Feynman gauge with three different IR regulators: nonzero gluon mass, nonzero quark mass and dimensional regularization (DR). 
Sec.~\ref{s:problem_matching} addresses matching for $e(x)$ and $h_{L}(x)$ in the $\msbar$ scheme. We summarize our results in Sec.~\ref{s:conclusions}. 

\section{Definitions}
\label{s:definitions}
We start by recalling the definition of twist-3 light-cone PDFs $e(x)$ and $h_L(x)$ for quarks. Generally, light-cone PDFs are defined through the correlation function\footnote{For a generic four-vector $v$ we
denote the Minkowski components by $(v^0, v^1, v^2, v^3)$ and
the light-cone components by $(v^+, v^-, \vec{v}_\perp)$, with
$v^+ = \frac{1}{\sqrt{2}} (v^0 + v^3)$, $v^- = \frac{1}{\sqrt{2}}
(v^0 - v^3)$ and $\vec{v}_\perp = (v^1, v^2)$.}
\begin{equation}
\Phi^{[\Gamma]}(x, S) = \frac{1}{2} \int \frac{dz^-}{2\pi} \, e^{i k \cdot z} \, \langle P, S | \bar{\psi}(- \tfrac{z}{2}) \, \Gamma \, {\cal W}(- \tfrac{z}{2}, \tfrac{z}{2}) \,\psi(\tfrac{z}{2})  | P, S \rangle \Big|_{z^+ = 0, \vec{z}_\perp = \vec{0}_\perp} \, .
\label{e:corr_standard_GPD}
\end{equation}
Here $\Gamma$ denotes a generic gamma matrix. Color gauge invariance of this bi-local quark-quark correlator is enforced by the Wilson line
\begin{equation}
{\cal W}(- \tfrac{z}{2}, \tfrac{z}{2}) \Big|_{z^+ = 0, \vec{z}_\perp = \vec{0}_\perp}
= {\cal P} \, \textrm{exp} \, \Bigg( - i g_s \int_{-\tfrac{z^-}{2}}^{\tfrac{z^-}{2}} \, dy^- \, A^+(0^+, y^-, \vec{0}_\perp) \Bigg) \,,
\end{equation}
where ${\cal P}$ is a path-ordered exponential depending on the plus-component of the gluon field. The hadron is characterized by its 4-momentum $P$ and a covariant spin vector $S$ which can be written as
\begin{equation}
S^{\mu} = (S^+, S^-, \vec{S}_\perp) = \bigg( \lambda \frac{P^+}{M}, - \lambda \frac{M}{2 P^+}, \vec{S}_\perp \bigg) \,,
\end{equation}
where $\lambda$ is the helicity of the hadron and $M$ is its mass. The spin vector satisfies the relation $P \cdot S = 0$ by definition. The twist-3 light-cone PDFs $e(x)$ and $h_L(x)$ are then defined as
\begin{eqnarray}
\Phi^{[1]} & = & \frac{1}{2P^+} \, \bar{u}(P,S) \, 1 \, u(P,S) \, e(x) = \frac{M}{P^+} \, e(x) \,,
\\[0.2cm]
\Phi^{[i \sigma^{+-} \gamma_5]} & = & \frac{1}{2P^+} \, \bar{u}(P,S) \, i \sigma^{+-} \gamma_5 \, u(P,S) \, h_L(x) = \frac{M}{P^+} \, \lambda \, h_L(x) \,,
\end{eqnarray}
where $u(P,S)$ ($\bar{u}(P,S)$) is the spinor for the incoming (outgoing) hadron, $\sigma^{\mu\nu}=\frac{i}{2}(\gamma^{\mu}\gamma^{\nu}-\gamma^{\nu}\gamma^{\mu})$ and $\gamma_{5}$ is the usual matrix which anti-commutes with any other Dirac matrix.

We now turn to the quasi-PDFs which are defined through the spatial correlation function~\cite{Ji:2013dva,Ji:2014gla}
\begin{equation}
\Phi_{\rm Q}^{[\Gamma]}(x, S; P^{3}) = \frac{1}{2} \int \frac{dz^3}{2\pi} \, e^{i k \cdot z} \, \langle P, S | \bar{\psi}(- \tfrac{z}{2}) \, \Gamma \, {\cal W}_{\rm Q}(- \tfrac{z}{2}, \tfrac{z}{2}) \,\psi(\tfrac{z}{2})  | P, S \rangle \Big|_{z^0 = 0, \vec{z}_\perp = \vec{0}_\perp} \,,
\label{e:corr_quasi_GPD}
\end{equation}
with the Wilson line
\begin{equation}
{\cal W}_{\rm Q}(- \tfrac{z}{2}, \tfrac{z}{2}) \Big|_{z^0 = 0, \vec{z}_\perp = \vec{0}_\perp}
= {\cal P} \, \textrm{exp} \, \Bigg( - i g_s \int_{-\tfrac{z^3}{2}}^{\tfrac{z^3}{2}} \, dy^3 \, A^3(0, \vec{0}_\perp, y^3) \Bigg) \,.
\end{equation}
The spin vector in this case is written as 
\begin{equation}
S^{\mu} = (S^{0}, \vec{S}_\perp, S^{3}) = \bigg( \lambda \frac{P^{3}}{M}, \vec{S}_\perp, \lambda \frac{P^{0}}{M} \bigg) \,.
\end{equation}
The quasi-PDFs of interest are then defined as
\begin{eqnarray}
\Phi_{\rm Q}^{[1]} = \frac{M}{P^{3}} \, e_{\rm Q}(x; P^{3}) \,, \qquad
\Phi_{\rm Q}^{[i \sigma^{30} \gamma_5]}  =  \frac{M}{P^{3}} \, \lambda \, h_{L, {\rm Q}}(x; P^{3}) \,.
\end{eqnarray} 
The definitions of the quasi-PDFs are such that their lowest moments are $P^{3}$ independent~\cite{Bhattacharya:2019cme},
\begin{eqnarray}
\int dx \, e_{\rm Q} (x; P^{3}) = \int dx \, e (x) \,, \qquad \int dx \, h_{L, \rm Q} (x; P^{3}) = \int dx \, h_{L} (x) \, .
\label{e:norm}
\end{eqnarray}

\section{One-Loop results}
\label{s:one_loop_results}
\begin{figure}[t]
\begin{center}
    \includegraphics[width=15cm]{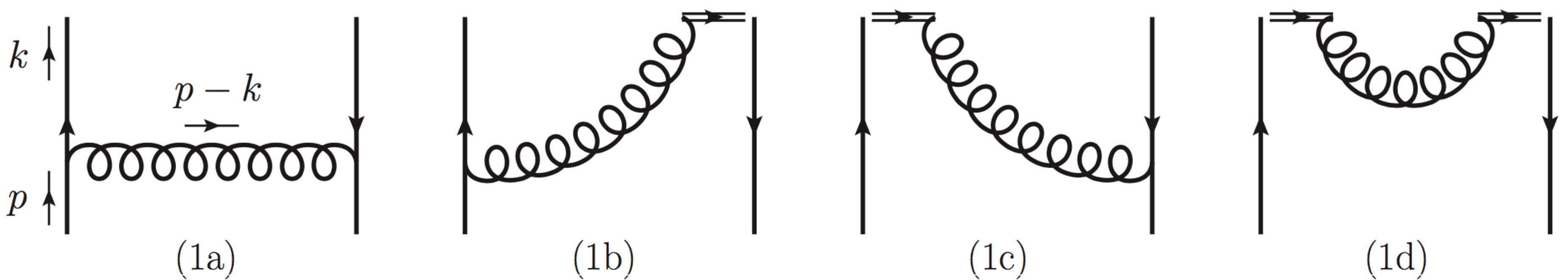}
 	\caption{One-loop real diagrams contributing to the light-cone PDFs $e(x)$ and $h_L(x)$, and the quasi-PDFs $e_{\rm Q}$ and $h_{L,\rm Q}$.}
 	\label{fig:real}
\end{center}
\end{figure}

\begin{figure}[t]
\begin{center}
\hspace{0.5cm}
    \includegraphics[width=14.6cm]{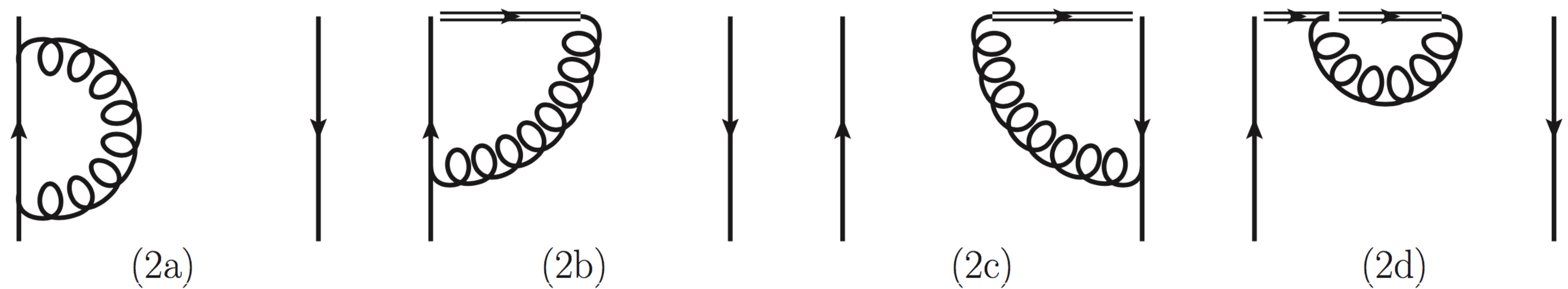}
 	\caption{One-loop virtual diagrams contributing to the light-cone PDFs $e(x)$ and $h_L(x)$, and the quasi-PDFs $e_{\rm Q}$ and $h_{L,\rm Q}$. The Hermitean conjugate diagrams of (2a) and (2d) have not been shown.}
 	\label{fig:virtual}
\end{center}
\end{figure}

In this section, we calculate the perturbative corrections to the light-cone PDFs and the quasi-PDFs to one-loop order. In principle, one can do these calculations in any gauge and the final result should be independent of the gauge. Here, we choose to work in the Feynman gauge for which the contributing real and virtual diagrams are shown in Fig.~\ref{fig:real} and Fig.~\ref{fig:virtual}, respectively. We regulate the infrared (IR) divergences by making use of 3 different schemes: non-zero parton mass regulations ($m_{g} \neq 0$ for gluon mass and $m_{q} \neq 0$ for quark mass) and dimensional regularization (DR). The ultraviolet (UV) divergences in the problem have consistently been tackled with DR. The individual diagrams have additional divergences at $x=1$. However, the combination of real and virtual corrections (which are proportional to $\delta(1-x)$) is well-defined. Since our computations are at the level of the partons (these results are prior to embedding them into a full correlator picture), we use $m_q$ and $p$ ($=x P$) as the mass and 4-momentum for the (quark) target.

\subsection{Results for $e(x)$}
\label{s:e}
In this subsection, we focus on the light-cone PDF $e(x)$ and its corresponding quasi-PDF $e_{\rm Q} (x)$.
\subsubsection{Light-cone PDF}
\label{s:light_cone_e}
Let us discuss first the computation of the real diagrams. The one-loop correction for Fig.~(1a) is calculated  as
\begin{eqnarray}
\dfrac{m_{q}}{p^{+}}\,e^{\rm{(1a)}} (x) = -\dfrac{i g^{2} C_{F} \mu^{2\epsilon} g_{\mu \nu}}{4} \int^{\infty}_{-\infty} \dfrac{d^{n}k}{(2\pi)^{n}} \dfrac{{\rm{Tr}} \big [ (\slashed{p}+ m_{q}) \, \gamma^{\nu} \, (\slashed{k}+m_{q}) \, 1 \, (\slashed{k}+m_{q}) \, \gamma^{\mu} \big ]}{(k^{2} -m^{2}_{q}+ i\varepsilon)^{2} ((p-k)^{2}-m^{2}_{g} + i\varepsilon)} \, \delta \bigg ( x- \dfrac{k^{+}}{p^{+}} \bigg ) \dfrac{1}{p^{+}} \, , 
\label{e:diagram_1a_LC}
\end{eqnarray}
where $g$ denotes the coupling for the quark-gluon-quark vertex and $C_F = 4/3$ is the color factor. The integrals in Eq.~(\ref{e:diagram_1a_LC}) have been analytically continued to $n=4-2\epsilon$ dimensions to regulate the divergences present otherwise. Here $\epsilon$ is the DR regulator. If $\epsilon$ is used for the UV divergences, then $\epsilon \rightarrow \epsilon_{\UV} > 0$ (and the corresponding subtraction scale is $\mu \rightarrow \mu_{\UV} > 0$), while if it is used for the IR divergences then $\epsilon \rightarrow \epsilon_{\IR} < 0$ (and $\mu \rightarrow \mu_{\IR} > 0$). Trace algebra simplifies Eq.~(\ref{e:diagram_1a_LC}) to
\begin{equation}
e^{\rm{(1a)}} (x) = -\dfrac{i g^{2} C_{F} \mu^{2\epsilon}}{(2\pi)^n} p^+ \int^{\infty}_{-\infty} d^{n-2}k_\perp d k^-  dk^+
\dfrac{(2-n) 2 \, p\cdot k + n (k^{2}+m^{2}_{q})} 
{(k^{2} -m^{2}_{q}+ i\varepsilon)^{2} ((p-k)^{2}-m^{2}_{g} + i\varepsilon)} \, \delta  \bigg ( x- \dfrac{k^{+}}{p^{+}} \bigg ) \dfrac{1}{p^{+}} \, .
\label{e:e_1a_LC_step1}
\end{equation}
We will use the following abbreviation to present our one-loop results
\begin{displaymath}
{\cal P}_{\UV} = \frac{1}{\epsilon_{\UV}} + \ln 4\pi - \gamma_E \,,
\end{displaymath}
and similarly ${\cal P}_{\IR}$ for the IR.
After regulating UV and IR divergences in the $k_{\perp}$ integrals, Eq.~(\ref{e:e_1a_LC_step1}) for $m_{g} \neq 0$ case can be written as
\begin{eqnarray}
e^{\rm{(1a)}} (x) \Big |_{m_{g}}  &=& e^{\rm{(1a)}}_{\rm{(s)}} (x) + e^{\rm{(1a)}}_{\rm{(c)}} (x) \Big |_{m_{g}}  \, ,
\end{eqnarray}
where the ``singular" part of the light-cone PDF $e(x)$ (denoted as $e_{(\rm s)}$) is given by
\begin{eqnarray}
e^{\rm{(1a)}}_{\rm{(s)}} (x) =
\begin{dcases}
& e^{\rm{(1a)}}_{\rm{(s)}} (x) \Big |_{m_{q}} = \dfrac{\alpha_{s} C_{F}}{2\pi} \, \delta (x) \bigg ( {\cal P}_{\UV} + \ln \dfrac{\mu^{2}_{\UV}}{m^{2}_{q}}-1 \bigg ) \, ,
\\[0.2cm]
& e^{\rm{(1a)}}_{\rm{(s)}} (x) \Big |_{\epsilon_{\IR}} = \dfrac{\alpha_{s} C_{F}}{2\pi} \, \delta (x) \bigg ( {\cal P}_{\UV} - {\cal P}_{\IR} + \ln \dfrac{\mu^{2}_{\UV}}{\mu^{2}_{\IR}} \bigg ) \, ,
\end{dcases}
\label{e:e_sing}
\end{eqnarray}
and the ``canonical" (or the regular) part of the light-cone PDF $e(x)$ (denoted as $e_{(\rm c)}$) is given by
\begin{eqnarray}
e^{\rm{(1a)}}_{\rm{(c)}} (x) \Big |_{m_{g}} & = & 
\dfrac{\alpha_{s} C_{F}}{2\pi} \bigg ( {\cal P}_{\UV} + \ln \dfrac{\mu^{2}_{\UV}}{x m^{2}_{g}} -  \dfrac{1-x}{x} \bigg ) \, .
\label{e:e_cano}
\end{eqnarray}
It is interesting to discuss the above results. We divided the result into two distinct parts: (a) singular, and (b) canonical. 
The singular part of the PDF has a zero-mode $\delta(x)$ contribution.
Such a singularity originates from a term proportional to $p \cdot k$ (see the first term in Eq.~(\ref{e:e_1a_LC_step1})), which can be used to cancel the gluon propagator leading to~\cite{Yan:1973qg, Aslan:2018tff}
\begin{eqnarray}
(2-n) \int dk^{-} \dfrac{1}{(k^{2}-m^{2}_{q}+ i\varepsilon)^{2}} = (2-n) \dfrac{i\pi}{(k^{2}_{\perp}+m^{2}_{q})} \dfrac{\delta(x)}{p^{+}} \, .
\label{e:zero_mode}
\end{eqnarray}
The $k_{\perp}$ integral in Eq.~(\ref{e:zero_mode}) has a UV divergence which is regulated by DR, and the coefficient of this integral is such that the UV pole $1/\epsilon_{\UV}$ allows for a $\delta(x)$ contribution in Eq.~(\ref{e:e_sing}). For $m_{g} \neq 0$, one should in principle set the quark mass term in Eq.~(\ref{e:zero_mode}) to zero. In doing so, we confront an IR divergence in the limit $k_{\perp} \rightarrow 0$. As we pointed out in Ref.~\cite{Bhattacharya:2020xlt}, this IR divergence is left unattended when one works with a nonzero gluon mass, and this is a new feature appearing at the level of twist-3. In fact, this insufficiency of the gluon mass as an IR regulator is only confined to this specific singular zero-mode term present in Fig.~(1a). For practical reasons, we suggest(ed) to handle the IR divergence by either retaining the quark mass term in Eq.~(\ref{e:zero_mode}) or by using DR. For $g_T$, the two methods lead to two (qualitatively) different answers, namely, the $\delta(x)$ drops out when using DR~\cite{Bhattacharya:2020xlt}. For $e(x)$, as well as for $h_L(x)$, the coefficient of the $k_{\perp}$ integral in Eq.~(\ref{e:zero_mode}) is such that, regardless of the IR scheme, the $\delta(x)$ term survives. There is another crucial difference between the $\delta(x)$ appearing here versus those in $g_T$. 
The $\delta (x)$ for $e(x)$ and $h_L(x)$ comes in with a prefactor that has an explicit dependence on the IR pole. On the other hand, the prefactor of $\delta (x)$ for $g_T$ is IR-finite. 
Note that the two results for the singular part of $e(x)$ in Eq.~(\ref{e:e_sing}) correspond to the two options of working with either $m_{q} \neq 0$, or DR for the $k_{\perp}$ integral in Eq.~(\ref{e:zero_mode}). For the canonical part of $e(x)$, $m_{g} \neq 0$ is sufficient to regulate the IR divergences and, therefore, we have a unique result in Eq.~(\ref{e:e_cano}).
With $m_{q} \neq 0$ and DR for the IR, one obtains
\begin{eqnarray}
e^{\rm{(1a)}} (x) \Big |_{m_{q}}  &=& e^{\rm{(1a)}}_{\rm{(s)}} (x) \Big |_{m_{q}} + e^{\rm{(1a)}}_{\rm{(c)}} (x) \Big |_{m_{q}} \nonumber \\[0.2cm]
&=& \dfrac{\alpha_{s} C_{F}}{2\pi} \, \delta (x) \bigg ( {\cal P}_{\UV} + \ln \dfrac{\mu^{2}_{\UV}}{m^{2}_{q}}-1 \bigg ) \nonumber + \dfrac{\alpha_{s} C_{F}}{2\pi} \bigg ( {\cal P}_{\UV} + \ln \dfrac{\mu^{2}_{\UV}}{(1-x)^{2}m^{2}_{q}} - \dfrac{2}{1-x} \bigg ) \, ,  \\[0.2cm]
e^{\rm{(1a)}} (x) \Big |_{\epsilon_{\IR}}  &=& e^{\rm{(1a)}}_{\rm{(s)}} (x) \Big |_{\epsilon_{\IR}} + e^{\rm{(1a)}}_{\rm{(c)}} (x) \Big |_{\epsilon_{\IR}} \nonumber \\[0.2cm]
&=& \dfrac{\alpha_{s} C_{F}}{2\pi} \, \delta (x) \bigg ( {\cal P}_{\UV} - {\cal P}_{\IR} + \ln \dfrac{\mu^{2}_{\UV}}{\mu^{2}_{\IR}} \bigg ) + \dfrac{\alpha_{s} C_{F}}{2\pi} \, \bigg (  {\cal P}_{\UV} - {\cal P}_{\IR} + \ln \dfrac{\mu^{2}_{\UV}}{\mu^{2}_{\IR}}  \bigg ) \, .
\end{eqnarray}
Therefore, for all three IR regulators, the $\delta(x)$ contributes.

The diagram of Fig.~(1b) is calculated as 
\begin{eqnarray}
\dfrac{m_{q}}{p^{+}}\,e^{\rm{(1b)}} (x) = -\dfrac{i g^{2} C_{F} \mu^{2\epsilon} g_{\mu \nu} v^{\nu}}{4} \int^{\infty}_{-\infty} \dfrac{d^{n}k}{(2\pi)^{n}} \dfrac{{\rm{Tr}} \big [ (\slashed{p}+ m_{q}) \, 1 \, (\slashed{k}+m_{q}) \, \gamma^{\mu} \big ]}{(v.(p-k)+i\varepsilon)(k^{2} -m^{2}_{q}+ i\varepsilon) ((p-k)^{2}-m^{2}_{g} + i\varepsilon)} \, \delta \bigg ( x- \dfrac{k^{+}}{p^{+}} \bigg ) \dfrac{1}{p^{+}} \, .
\label{e:diagram_1b_LC}
\end{eqnarray}
Here, $v$ is defined such that $v^{2} =0$ and $v \cdot a = a^{+}$ for any four-vector $a^{\mu}$. The results for the three IR regulators are
\begin{eqnarray}
e^{\rm{(1b)}}(x) \Big |_{m_{g}} & = & \frac{\alpha_s C_F}{2\pi} \, \frac{1 + x}{2(1-x)} \, \bigg ( {\cal P}_{\UV} + \ln \frac{\mu_{\UV}^2}{x m_g^2} \bigg ) \, , \\[0.2cm]
e^{\rm{(1b)}}(x) \Big |_{m_{q}} & = & \frac{\alpha_s C_F}{2\pi} \, \frac{1 + x}{2(1-x)} \, \bigg ( {\cal P}_{\UV} + \ln \frac{\mu_{\UV}^2}{(1-x)^2 m_q^2} \bigg ) \, , \\[0.2cm]
e^{\rm{(1b)}}(x) \Big |_{\epsilon_{\IR}} & = & \frac{\alpha_s C_F}{2\pi} \, \frac{1 + x}{2(1-x)} \, \bigg ( {\cal P}_{\UV} - {\cal P}_{\IR} + \ln \frac{\mu_{\UV}^2}{\mu_{\IR}^2} \bigg ) \, .
\end{eqnarray}
The diagram of Fig.~(1c) gives the same result as the one of Fig.~(1b). For the light-cone PDFs, the diagram of Fig.~(1d) drops out because the results are proportional to $v^{2}$.

We now proceed with the computation of the virtual diagrams. The quark self-energy diagram in Fig.~(2a) is independent of the Dirac structure and we presented it in Ref.~\cite{Bhattacharya:2020xlt}. We quote the results here for the sake completeness,
\begin{eqnarray}
e^{\rm{(2a)}} (x) \Big |_{m_{g}} &=& 
-\dfrac{\alpha_{s} C_{F}}{2 \pi} \int_{0}^{1} dy \, y  \bigg ( {\cal P}_{\UV} + \ln \dfrac{\mu_{\UV}^{2}}{y m^{2}_{g} } -1 \bigg ) \, , \\[0.2cm]
e^{\rm{(2a)}} (x) \Big |_{m_{q}} &=& 
-\dfrac{\alpha_{s} C_{F}}{2 \pi} \int_{0}^{1} dy \, (1-y)  \bigg ( {\cal P}_{\UV} + \ln \dfrac{\mu_{\UV}^{2}}{(1-y)^2 m_q^{2}} - \dfrac{1+y^2}{(1-y)^{2}} \bigg ) \, , \\[0.2cm]
e^{\rm{(2a)}} (x) \Big |_{\epsilon_{\IR}} &=& 
-\dfrac{\alpha_{s} C_{F}}{2 \pi} \int_{0}^{1} dy \, y  \bigg ( {\cal P}_{\UV} - {\cal P}_{\IR}  + \ln \dfrac{\mu_{\UV}^{2}}{\mu_{\IR}^{2} } \bigg ) \, ,
\label{e:diagram_2a_LC}
\end{eqnarray}
where $y$ is the (integrated) loop momentum fraction.

The initial expression for the diagrams of Fig.~(2b) and Fig.~(2c), is the same as the ones of Fig.~(1b) and Fig.~(1c), respectively, modulo an overall sign (see Ref.~\cite{Bhattacharya:2020xlt}). Therefore the results are
\begin{eqnarray}
e^{\rm{(2b)}} (x) \Big |_{m_{g}} & = & - \frac{\alpha_s C_F}{2\pi} \int_{0}^{1} dy \, \frac{1 + y}{2(1-y)} \, \bigg ( {\cal P}_{\UV} + \ln \frac{\mu_{\UV}^2}{y m_g^2} \bigg ) \, , \\[0.2cm]
e^{\rm{(2b)}} (x) \Big |_{m_{q}} & = & - \frac{\alpha_s C_F}{2\pi} \int_{0}^{1} dy \, \frac{1 + y}{2(1-y)} \, \bigg ( {\cal P}_{\UV} + \ln \frac{\mu_{\UV}^2}{(1-y)^2 m_q^2} \bigg ) \, , \\[0.2cm]
e^{\rm{(2b)}} (x) \Big |_{\epsilon_{\IR}} & = & - \frac{\alpha_s C_F}{2\pi} \int_{0}^{1} dy \, \, \frac{1 + y}{2(1-y)} \, \bigg ( {\cal P}_{\UV} - {\cal P}_{\IR} + \ln \frac{\mu_{\UV}^2}{\mu_{\IR}^2} \bigg ) \, .
\end{eqnarray}
Finally, the diagram of Fig.~(2d) does not contribute, similar to the corresponding real diagram of Fig.~(1d). All these results for the virtual diagrams are to be understood with an overall prefactor of $\delta(1-x)$ which we have left out for simplicity.

\subsubsection{Quasi-PDF}
Just as in the light-cone case, the result for Fig.~(1a) can be divided into a singular and a canonical part. In this section, we focus mostly on discussing the subtleties involved in the treatment/calculation of the singular part for the quasi-PDFs, and we refer to Ref.~\cite{Bhattacharya:2020xlt} for more details on the calculation of the canonical part and the other diagrams. The term which generates a $\delta(x)$ in the light-cone PDF $e(x)$ gives rise to the following structure for the quasi-PDF $e_{\rm Q}(x)$\footnote{For convenience of notation, in our results we use that $p^{2}_{3} = (p^{3})^{2}$.}:
\begin{eqnarray}
e^{\rm{(1a)}}_{\rm{Q(s)}} (x)=
\begin{dcases}
& e^{\rm{(1a)}}_{\rm{Q(s)}} (x) \Big |_{m_{q}}  =  \dfrac{\alpha_{s} C_{F}}{2\pi}
\begin{dcases}
\phantom{+} \dfrac{1}{x}
& \quad x > 1  \\[0.2cm]
\phantom{+} \dfrac{p^{3}}{\sqrt{x^{2}p^{2}_{3} + m^{2}_{q}}}
& \quad -1 < x < 1 \\[0.2cm]
- \dfrac{1}{x}
& \quad x < -1 \, , 
\end{dcases}
\\[0.3cm]
& e^{\rm{(1a)}}_{\rm{Q(s)}} (x) \Big |_{\epsilon_{\IR}}  =  \dfrac{\alpha_{s} C_{F}}{2\pi}
\begin{dcases}
\phantom{+} \dfrac{1}{x}
& \quad x > 1  \\[0.2cm]
\phantom{+} C(\epsilon_{\IR}) \bigg (\frac{p^3}{\mu_{\IR}} \bigg )^{-2\epsilon_{\IR}} \dfrac{(1-\epsilon_{\IR})}{|x|^{1+2\epsilon_{\IR}}}
& \quad -1 < x < 1 \\[0.2cm]
- \dfrac{1}{x}
& \quad x < -1 \, ,
\end{dcases}  
\end{dcases}
\label{e:unexpanded_eq_mg}
\end{eqnarray}
where
\begin{equation}
C(\epsilon_{\IR})=\dfrac{\pi^{1/2-\epsilon_{\IR}}}{(2\pi)^{-2\epsilon_{\IR}}
\Gamma[1/2-\epsilon_{\IR}]} \, .
\label{e:def_C_angular}
\end{equation}
Note that we call the terms in Eq.~(\ref{e:unexpanded_eq_mg}) ``singular" because, as we shall see in the following, they exhibit an IR singularity at $x = 0$. However, these terms are well-behaved in other regions of $x$.
Recall that the gluon mass does not enter into the calculation of the singular components of the PDFs. DR for the $k^{3}$ component in $1-2\epsilon$ dimensions leads to extra factors like $C(\epsilon_{\IR})$ as shown in Eq.~(\ref{e:def_C_angular}). The $x$-dependent results for the real diagrams for the quasi-PDFs are UV finite. (The \textit{same} UV divergences which stem from the $k_{\perp}$ integrals for the light-cone PDFs show up in the $x$-\textit{integrals} of the above results for the quasi-PDFs.) Therefore, at the level of extracting the $x$-dependence, one can already perform a Taylor expansion in powers of the UV regulator, that is $\epsilon_{\UV} \approx 0$ for the regions $|x| > 1 $. 
One can do this expansion in powers of the IR regulator, that is $m^{2}_{q}/p^{2}_{3} \approx 0$ or $\epsilon_{\IR} \approx 0$, for instance for the region $-1 < x < 1$, but one \textit{cannot} do so at the specific point of $x = 0$. If we would do that, we would arrive at the result $e_{\rm{Q(s)}} \thicksim (1/x)$.
This generates a misleading result and one arrives at the incorrect conclusion that the functional forms for the IR singularities do not agree between the light-cone $e_{\rm{(s)}}$ and quasi-PDF $e_{\rm{Q(s)}}$. This is because for the light-cone result one has the IR singularity associated with the delta function $e_{\rm(s)} \thicksim \delta(x) \ln m^{2}_{q}$ or $e_{\rm(s)} \thicksim \delta(x) 1/\epsilon_{\IR}$, while for the quasi-PDF the IR singularity is reflected in the structure $e_{\rm{Q(s)}} \thicksim (1/x)$ as $x \rightarrow 0$. A conceptually correct approach is to hold off with this Taylor expansion in powers of the IR regulator for the singular terms until we have carefully isolated the IR singularity at $x=0$. 

We first take up the $m_{q} \neq 0$ case. In the following, we integrate $\dfrac{1}{\sqrt{x^{2} + \eta^{2}}}$, with $\eta \equiv m_{q}/p^{3}$, over $x$ in the interval $[-1, \, 1]$ with a test function $f(x)$ that is finite at the origin,
\begin{eqnarray}
\int^{1}_{-1} dx \, \dfrac{f(x)}{\sqrt{x^{2} + \eta^{2}}} &=& \int^{0}_{-1} dx \dfrac{(f(x) - f(0))}{\sqrt{x^{2} + \eta^{2}}} + \int^{1}_{0} dx \dfrac{(f(x) - f(0))}{\sqrt{x^{2} + \eta^{2}}} + \int^{1}_{-1} dx \, \dfrac{f(0)}{\sqrt{x^{2} + \eta^{2}}}
\label{e:distribution_mq_step1} \, .
\end{eqnarray}
For the first and the second terms, we perform an expansion about $\eta \approx 0$. For the third term, we first integrate and then expand the result about $\eta \approx 0$. Doing these steps, we arrive at
\begin{eqnarray}
\int^{1}_{-1} dx \, \dfrac{f(x)}{\sqrt{x^{2} + \eta^{2}}} &=& \int^{0}_{-1} dx \dfrac{(f(x) - f(0))}{- x}  + \int^{1}_{0} dx \dfrac{(f(x) - f(0))}{x}  + f(0) \bigg ( \ln \dfrac{4}{\eta^{2}} \bigg ) + \mathcal{O} (\eta^{2})  \, . \qquad
\label{e:distribution_mq_step2} 
\end{eqnarray}
We now make the identifications
\begin{eqnarray}
\int^{0}_{-1} dx \, \dfrac{(f(x) - f(0))}{- x} &=& \int^{0}_{-1} dx \, f(x)\, \bigg [ \dfrac{1}{-x} \bigg ]_{+[0]} \, ,\\[0.2cm]
\label{e:plus_mqneg}
\int^{1}_{0} dx \, \dfrac{(f(x) - f(0))}{x} &=& \int^{1}_{0} dx \, f(x) \, \bigg [ \dfrac{1}{x} \bigg ]_{+[0]} \, ,
\label{e:plus_mqpos}
\end{eqnarray}
based on the definition of plus-functions at $x =0$, that is,
\begin{eqnarray}
{\rm{R}}_{0}(|x|) &\equiv& \bigg [ \dfrac{1}{|x|} \bigg ]_{+[0]}  = \theta(|x|) \, \theta(1-|x|) \lim_{\beta \rightarrow 0} \bigg [ \dfrac{\theta(|x| - \beta)}{|x|} + \delta(|x| - \beta) \ln \beta \bigg ] \, ,
\label{e:R0}
\end{eqnarray}
where $-1 < x < 1$ and $\beta >0$.
(Note that the right-hand side of Eq.~(\ref{e:R0}) is an exact mathematical way of computing plus-functions at $x = 0$ whose general definition has been shown in Eq.~(\ref{e:def_plus_0}). In this context, we also refer to Eq.~(\ref{e:our_def}) and the paragraph thereafter.)
Making these replacements, we see that Eq.~(\ref{e:distribution_mq_step2}) can be cast as 
\begin{eqnarray}
\int^{1}_{-1} dx \, \dfrac{f(x)}{\sqrt{x^{2} + \eta^{2}}} &=& \int^{1}_{-1} dx \, f(x) \, \bigg [ \dfrac{1}{|x|} \bigg ]_{+[0]}  + \int^{1}_{-1} dx \, f(x) \, \delta(x) \bigg ( \ln \dfrac{4}{\eta^{2}} \bigg ) + \mathcal{O} (\eta^{2} ) \, .
\label{e:distribution_mq_step3}
\end{eqnarray}
Since Eq.~(\ref{e:distribution_mq_step3}) holds for any arbitrary test function in the interval $[-1, \, 1]$, we arrive at the identity 
\begin{eqnarray}
\dfrac{\theta(1-x) \, \theta(1+x)}{\sqrt{x^{2} + \eta^{2}}} = \delta(x) \bigg ( \ln \dfrac{4}{\eta^{2}} \bigg ) + {\rm{R}}_{0}(|x|) + \mathcal{O} (\eta^{2} ) \, .
\label{e:plus_mq}
\end{eqnarray}

We can extend the same logic for the case of DR. One can readily verify the identity
\begin{eqnarray}
\dfrac{\theta(1-x) \, \theta(1+x)}{|x|^{1+2\epsilon_{\IR}}} = -\dfrac{\delta(x)}{\epsilon_{\IR}} + \sum^{\infty}_{n=0} (-1)^{n} \dfrac{(2\epsilon_{\IR})^{n}}{n!} \, {\rm{R}}_{n}(|x|) \, ,
\label{e:plus_DR}
\end{eqnarray}
where
\begin{eqnarray}
{\rm{R}}_{n}(|x|) \equiv \bigg [ \dfrac{\ln^{n} |x|}{|x|} \bigg ]_{+[0]} = \theta(|x|) \, \theta(1-|x|) \lim_{\beta \rightarrow 0} \bigg [ \theta(|x| - \beta) \dfrac{\ln^{n} |x|}{|x|} + \delta(|x| - \beta) \dfrac{\ln^{n+1}\beta}{n+1} \bigg ] \, ,
\label{e:general_Rn}
\end{eqnarray}
with $\beta >0$. This point has been addressed in Ref.~\cite{Izubuchi:2018srq} for $x > 0$. The expression in Eq.~(\ref{e:general_Rn}) is therefore a straightforward generalization of the relevant identity (see Eq.~(C.2)) of Ref.~\cite{Izubuchi:2018srq} in order to cover also the region of negative $x$. A simple way of understanding Eq.~(\ref{e:general_Rn}) is through the standard definition of plus-functions. A plus-function at $x = 0$, for instance in the interval $[0, \, 1]$, is defined as
\begin{eqnarray}
[f(x)]_{+[0]} \equiv \theta(x) \, \theta (1-x) \lim_{\beta \rightarrow 0} \bigg [ \theta(x-\beta) \, f(x) - \delta (x - \beta) \int^{1}_{\beta} dy \, f(y) \bigg ] \, .
\label{e:def_plus_0}
\end{eqnarray}
Therefore, for the specific case when the function is $1/x^{1+\epsilon}$, Eq.~(\ref{e:def_plus_0}) right away provides
\begin{eqnarray}
\bigg [ \dfrac{\theta(x) \, \theta(1-x)}{x^{1+2\epsilon}} \bigg ]_{+[0]} = \theta(x) \, \theta (1-x) \lim_{\beta \rightarrow 0} \bigg [ \dfrac{\theta(x - \beta)}{x^{1+2\epsilon}} - \delta (x - \beta) \bigg ( \dfrac{-1 + \beta ^{-2\epsilon}}{2\epsilon} \bigg ) \bigg ] \, .
\label{e:our_def}
\end{eqnarray}
A small-$\epsilon$ expansion of Eq.~(\ref{e:our_def}) readily shows its equivalence with Eq.~(\ref{e:general_Rn}). We refer to Ref.~\cite{Ligeti:2008ac} where general relations like in Eq.~(\ref{e:our_def}) were derived for different $x$-intervals, and various other important properties of plus-functions were outlined. For our purpose, the first two non-trivial terms in the expansion of $\epsilon_{\IR}$ matter in Eq.~(\ref{e:plus_DR}). Notice that the finite terms entering $\mathcal{O}(\eta^{0})$ and $\mathcal{O}(\epsilon^{0}_{\IR})$ are exactly the same for the two IR regulators.

Now, for $m_{g} \neq 0$, the result for $e_{\rm{Q}}$ can be written as
\begin{eqnarray}
e^{\rm{(1a)}}_{\rm{Q}} (x) \Big |_{m_{g}}  &=& e^{\rm{(1a)}}_{\rm{Q(s)}} (x) + e^{\rm{(1a)}}_{\rm{Q(c)}} (x) \Big |_{m_{g}}  \, .
\end{eqnarray}
Using Eq.~(\ref{e:plus_mq}) and Eq.~(\ref{e:plus_DR}) in Eq.~(\ref{e:unexpanded_eq_mg}), we obtain the following expressions for the singular terms for $e_{\rm{Q}}$
\begin{eqnarray}
e^{\rm{(1a)}}_{\rm{Q(s)}} (x) =
\begin{dcases}
& e^{\rm{(1a)}}_{\rm{Q(s)}} (x) \Big |_{m_{q}}  =  \dfrac{\alpha_{s} C_{F}}{2\pi}
\begin{dcases}
\phantom{+} \dfrac{1}{x}
& \quad x > 1  \\[0.2cm]
\phantom{+} \delta(x) \ln \dfrac{4p^{2}_{3}}{m^{2}_{q}} \, + \, {\rm{R}}_{0}(|x|) 
& \quad -1 < x < 1 \\[0.2cm]
- \dfrac{1}{x}
& \quad x < -1 \, , 
\end{dcases}
\\[0.3cm]
& e^{\rm{(1a)}}_{\rm{Q(s)}} (x) \Big |_{\epsilon_{\IR}}  =  \dfrac{\alpha_{s} C_{F}}{2\pi}
\begin{dcases}
\phantom{+} \dfrac{1}{x}
& \quad x > 1  \\[0.2cm]
\phantom{+} - \delta(x) \bigg ( {\cal P}_{\IR} - 1 - \ln \dfrac{4p^{2}_{3}}{\mu^{2}_{\IR}} \bigg ) \, + \, {\rm{R}}_{0}(|x|) 
& \quad -1 < x < 1 \\[0.2cm]
- \dfrac{1}{x}
& \quad x < -1 \, .
\end{dcases}  
\end{dcases}
\end{eqnarray}
As discussed in Sec.~\ref{s:light_cone_e}, there are two results because $m_{g} \neq 0$ is insufficient to regulate the IR divergence present in the singular terms. It is straightforward to arrive at the following result for the canonical part of $e_{\rm{Q}}$ with $m_{g} \neq 0$:
\begin{eqnarray}
e_{\rm{Q(c)}}^{\rm{(1a)}} (x) \Big |_{m_{g}}  & = & \frac{\alpha_s C_F}{2\pi}
\begin{dcases}
\ln \frac{x}{x-1} 
& \quad  x > 1  \\[0.2cm]
\ln \dfrac{4(1-x)p^{2}_{3}}{m^{2}_{g}} - \dfrac{1-x}{x}
& \quad  0 < x < 1 \\[0.2cm]
\ln \frac{x - 1}{x}
& \quad  x < 0 \, .
\label{e:quasi_e_mg}
\end{dcases}
\end{eqnarray}
For $m_{q} \neq 0$ and DR, we find
\begin{eqnarray}
e_{\rm{Q}}^{\rm{(1a)}} (x) \Big |_{m_{q}} & = & e_{\rm{Q(s)}}^{\rm{(1a)}} (x) \Big |_{m_{q}}  + e_{\rm{Q(c)}}^{\rm{(1a)}} (x) \Big |_{m_{q}} \nonumber \\[0.2cm] 
& = & \frac{\alpha_s C_F}{2\pi}
\begin{dcases}
\phantom{+} \dfrac{1}{x}
&  x > 1  \\[0.2cm]
\phantom{+} \delta(x) \ln \dfrac{4p^{2}_{3}}{m^{2}_{q}} \, + \, {\rm{R}}_{0}(|x|) 
& -1 < x < 1 \\[0.2cm]
- \dfrac{1}{x}
&  x < -1  
\end{dcases}
\, + \,
\begin{dcases}
\ln \frac{x}{x-1} 
&  x > 1  \\[0.2cm]
\ln \dfrac{4x p^{2}_{3}}{(1-x)m^{2}_{q}} - \dfrac{2}{1-x}
&  0 < x < 1 \\[0.2cm]
\ln \frac{x - 1}{x}
&  x < 0 \, , 
\end{dcases} 
\\[0.24cm]
e_{\rm{Q}}^{\rm{(1a)}} (x) \Big |_{\epsilon_{\IR}} & = & e_{\rm{Q(s)}}^{\rm{(1a)}} (x) \Big |_{\epsilon_{\IR}} + e_{\rm{Q(c)}}^{\rm{(1a)}} (x) \Big |_{\epsilon_{\IR}} \nonumber \\[0.2cm] 
& = & \frac{\alpha_s C_F}{2\pi}
\begin{dcases}
\phantom{+} \dfrac{1}{x}
&  x > 1  \\[0.2cm]
\phantom{+} - \delta(x) \bigg ( {\cal P}_{\IR} - 1 - \ln \dfrac{4p^{2}_{3}}{\mu^{2}_{\IR}} \bigg ) \, + \, {\rm{R}}_{0}(|x|) 
& -1 < x < 1 \\[0.2cm]
- \dfrac{1}{x}
&  x < -1 
\end{dcases}
\, + \,
\begin{dcases}
\ln \frac{x}{x-1}
&  x > 1  \\[0.2cm]
- {\cal P}_{\IR} + \ln \dfrac{4x(1-x)p^{2}_{3}}{\mu^{2}_{\IR}}
&  0 < x < 1 \qquad \\[0.2cm]
\ln \frac{x - 1}{x}
&  x < 0  \, .
\end{dcases}
\end{eqnarray}
Finally, we want to emphasize that Eq.~(\ref{e:plus_mq}) and Eq.~(\ref{e:plus_DR}) are to be understood in the sense of a distribution because $\delta (x)$ and $[...]_{+[0]}$ have a meaning under integrals only. In this sense, the singular terms $1/\sqrt{x^{2}+\eta^{2}}$ and $1/|x|^{1+2\epsilon_{\IR}}$ present in the quasi-PDF $e_{\rm{Q}}$ can be re-written so that, to leading order in $\mathcal{O}(\eta^{2})$ and $\mathcal{O}(\epsilon_{\IR})$, they reproduce exactly the same effect as the terms $\delta(x) \ln \eta^{2}$ or $\delta(x) 1/\epsilon_{\IR}$ present in the light-cone PDF $e_{\rm{(s)}}$. With these results, one infers an exact agreement in the IR-pole structures between the light-cone $e_{\rm(s)}$ $(e_{\rm(c)})$ and the quasi-PDF $e_{\rm{Q(s)}}$ ($e_{\rm{Q(c)}}$).

The contribution from the diagram of Fig.~(1b) is given by
\begin{eqnarray}
e_{\rm{Q}}^{\rm{(1b)}} (x) \Big |_{m_{g}} & = & \frac{\alpha_s C_F}{2\pi} \, \frac{1 + x}{2 (1 - x)}
\begin{dcases}
\ln \frac{x}{x - 1}
& \quad x > 1  \\[0.2cm]
\ln \frac{4(1 - x) p^{2}_{3}}{m_g^2}
& \quad 0 < x < 1 \\[0.2cm]
\ln \frac{x - 1}{x}
& \quad x < 0 \, ,
\end{dcases} 
\\[0.2cm]
e_{\rm{Q}}^{\rm{(1b)}} (x) \Big |_{m_{q}} & = & \frac{\alpha_s C_F}{2\pi} \, \frac{1 + x}{2 (1 - x)}
\begin{dcases}
\ln \frac{x}{x - 1}
& \qquad x > 1  \\[0.2cm]
\ln \frac{4x p^{2}_{3}}{(1-x) m_q^2}
& \qquad 0 < x < 1 \\[0.2cm]
\ln \frac{x - 1}{x}
& \qquad x < 0 \, ,
\end{dcases} 
\\[0.2cm]
e_{\rm{Q}}^{\rm{(1b)}} (x) \Big |_{\epsilon_{\IR}} & = & \frac{\alpha_s C_F}{2\pi} \, \frac{1 + x}{2 (1 - x)}
\begin{dcases}
\ln \frac{x}{x - 1}
& \quad x > 1  \\[0.2cm]
-{\cal P}_{\IR} + \ln \frac{4x(1-x) p^{2}_{3}}{\mu_{\IR}^2}
& \quad 0 < x < 1 \\[0.2cm]
\ln \frac{x - 1}{x}
& \quad x < 0 \, .
\end{dcases}
\end{eqnarray}
The diagram of Fig.~(1c) gives the same result as above. Unlike the case of light-cone PDFs, the diagram of Fig.~(1d) is non-vanishing for the quasi-PDFs and the result is given by
\begin{eqnarray}
e_{\rm{Q}}^{\rm{(1d)}} (x) & = & \frac{\alpha_s C_F}{2\pi}
\begin{dcases}
\frac{1}{1 - x}
& \quad x > 1  \\[0.2cm]
\frac{1}{x - 1}
& \quad 0 < x < 1 \\[0.2cm]
\frac{1}{x - 1}
& \quad x < 0 \, ,
\end{dcases}
\end{eqnarray} 
with the result being independent of the IR regulator.

We now take up the virtual diagrams. The quark self-energy diagram, which has been computed in our previous work~\cite{Bhattacharya:2020xlt}, is given by
\begin{eqnarray}
e^{\rm{(2a)}}_{\rm Q} (x) \Big |_{m_{g}} & = & - \dfrac{\alpha_{s} C_{F}}{2\pi}(1-\epsilon_{\UV}) C(\epsilon_{\UV})
\bigg (\frac{p^3}{\mu_{\UV}} \bigg )^{-2\epsilon_{\UV}} \int dy
\begin{dcases}
y^{\boldsymbol{-} 2\epsilon_{\UV}} \bigg ( y \ln \dfrac{y}{y-1} - 1\bigg )
&  y > 1  \\[0.2cm]
y^{\boldsymbol{-} 2\epsilon_{\UV}} \bigg ( y \ln \dfrac{4(1-y)p^{2}_{3}}{m_g^2}  + 1 -2y \bigg )
&  0 < y < 1 \quad \\[0.2cm]
(-y)^{\boldsymbol{-} 2\epsilon_{\UV}} \bigg ( y \ln \dfrac{y-1}{y} + 1 \bigg )
&  y < 0 \,,
\end{dcases} 
\end{eqnarray}
\begin{eqnarray}
e^{\rm{(2a)}}_{\rm Q} (x) \Big |_{m_{q}} & = & -\dfrac{\alpha_{s} C_{F}}{2\pi} C(\epsilon_{\UV})
\bigg (\frac{p^3}{\mu_{\UV}} \bigg )^{-2\epsilon_{\UV}} \int dy
\begin{dcases}
(1-\epsilon_{\UV}) \, y^{\boldsymbol{-} 2\epsilon_{\UV}} \bigg ((1-y) \ln \dfrac{y}{y-1} + 1\bigg )
& \, y > 1  \\[0.2cm]
y^{\boldsymbol{-} 2\epsilon_{\UV}} \bigg ( (1-\epsilon_{\UV}) (1 - y) \ln \dfrac{4 y p^{2}_{3}}{(1-y) m_q^2} \\[0.2cm]
- (1-\epsilon_{\UV}) \frac{2 y^2 - 5 y + 1}{1-y} 
-\bigg (1-\dfrac{\epsilon_{\UV}}{2} \bigg ) \frac{4y}{1-y} \bigg )
& \, 0 < y < 1 \\[0.2cm]
(1-\epsilon_{\UV}) \, (-y)^{\boldsymbol{-} 2\epsilon_{\UV}} \bigg ((1-y) \ln \dfrac{y-1}{y} - 1 \bigg )
& \, y < 0 \,,
\end{dcases} 
\\[0.2cm]
e^{\rm{(2a)}}_{\rm Q} (x) \Big |_{\epsilon_{\IR}} & = & -\dfrac{\alpha_{s} C_{F}}{2\pi}(1-\epsilon_{\UV}) C(\epsilon_{\UV})
\bigg (\frac{p^3}{\mu_{\UV}} \bigg )^{-2\epsilon_{\UV}} \int dy
\begin{dcases}
y^{\boldsymbol{-} 2\epsilon_{\UV}} \bigg ( y \ln \dfrac{y}{y-1} - 1\bigg )
& \, y > 1  \\[0.2cm]
y^{\boldsymbol{-} 2\epsilon_{\UV}} \bigg (- y \, {\cal P}_{\IR} + y \ln \dfrac{4y(1-y) p^{2}_{3}}{\mu_{\IR}^2} \\[0.2cm]
+ \, 1 - y \bigg )
& \, 0 < y < 1  \\[0.2cm]
(-y)^{\boldsymbol{-} 2\epsilon_{\UV}} \bigg ( y \ln \dfrac{y-1}{y} + 1 \bigg )
& \, y < 0 \,,
\end{dcases}
\end{eqnarray}
where $C(\epsilon_{\UV})$ is the same as in Eq.~(\ref{e:def_C_angular}), but with the replacement $\epsilon_{\IR} \rightarrow \epsilon_{\UV}$ and with the understanding that $\epsilon_{\UV} > 0$.
The (integrated) loop momentum fraction $y$ is defined through the relation $k^{3} = y p^{3}$. A detailed discussion of these self-energy results can be found in Ref.~\cite{Bhattacharya:2020xlt}.

For the diagram of Fig.~(2b) we find
\begin{eqnarray}
e^{\rm{(2b)}}_{\rm Q} (x) \Big |_{m_{g}} & = & -\dfrac{\alpha_{s} C_{F}}{2\pi}
C(\epsilon_{\UV}) \bigg (\frac{p^3}{\mu_{\UV}} \bigg )^{-2\epsilon_{\UV}}
\int dy \frac{1+y}{2(1-y)}
\begin{dcases}
y^{\boldsymbol{-} 2\epsilon_{\UV}} \ln \frac{y}{y - 1}
& \quad y > 1  \\[0.2cm]
y^{\boldsymbol{-} 2\epsilon_{\UV}} \ln \frac{4(1 - y) p^{2}_{3}}{m_g^2}
& \quad 0 < y < 1 \\[0.2cm]
(-y)^{\boldsymbol{-} 2\epsilon_{\UV}} \ln \frac{y - 1}{y}
& \quad y < 0 \,,
\end{dcases} 
\\[0.2cm]
e^{\rm{(2b)}}_{\rm Q} (x) \Big |_{m_{q}} & = & -\dfrac{\alpha_{s} C_{F}}{2\pi}
C(\epsilon_{\UV}) \bigg (\frac{p^3}{\mu_{\UV}} \bigg )^{-2\epsilon_{\UV}}
\int dy \frac{1+y}{2(1-y)}
\begin{dcases}
y^{\boldsymbol{-} 2\epsilon_{\UV}} \ln \frac{y}{y - 1}
& \quad y > 1  \\[0.2cm]
y^{\boldsymbol{-} 2\epsilon_{\UV}} \ln \frac{4 y p^{2}_{3}}{(1-y)m_q^2}
& \quad 0 < y < 1 \\[0.2cm]
(-y)^{\boldsymbol{-} 2\epsilon_{\UV}} \ln \frac{y - 1}{y}
& \quad y < 0 \,,
\end{dcases} 
\\[0.2cm]
e^{\rm{(2b)}}_{\rm Q} (x) \Big |_{\epsilon_{\IR}} & = &  -\dfrac{\alpha_{s} C_{F}}{2\pi}
C(\epsilon_{\UV}) \bigg (\frac{p^3}{\mu_{\UV}} \bigg )^{-2\epsilon_{\UV}}
\int dy \frac{1+y}{2(1-y)}
\begin{dcases}
y^{\boldsymbol{-} 2\epsilon_{\UV}} \ln \frac{y}{y - 1}
&  y > 1  \\[0.2cm]
y^{\boldsymbol{-} 2\epsilon_{\UV}} \bigg (- {\cal P}_{\IR} + \ln \frac{4y(1 - y) p^{2}_{3}}{\mu_{\IR}^2} \bigg )
&  0 < y < 1 \\[0.2cm]
(-y)^{\boldsymbol{-} 2\epsilon_{\UV}} \ln \frac{y - 1}{y}
&  y < 0 \,,
\end{dcases}
\end{eqnarray}
and the diagram in Fig.~(2c) gives the exact same result.

Finally, we find the following for the diagram in Fig.~(2d) 
\begin{eqnarray}
e^{\rm{(2d)}}_{\rm Q} (x) =  -\dfrac{\alpha_{s} C_{F}}{2\pi}
C(\epsilon_{\UV}) \bigg (\frac{p^3}{\mu_{\UV}} \bigg )^{-2\epsilon_{\UV}}
\int dy 
\begin{dcases}
y^{\boldsymbol{-} 2\epsilon_{\UV}}  \frac{1}{1-y}
& \quad y > 1  \\[0.2cm]
y^{\boldsymbol{-} 2\epsilon_{\UV}} \frac{1}{y-1}
& \quad 0 < y < 1 \\[0.2cm]
(-y)^{\boldsymbol{-} 2\epsilon_{\UV}} \frac{1}{y-1}
& \quad y < 0 \, .
\end{dcases} 
\end{eqnarray}
All of the $y$ integrals appearing in the virtual diagrams are logarithmically divergent. These UV divergences can be renormalized in the $\msbar$ scheme.

\subsection{Results for $h_{L}$}
\label{s:h_L}
In this subsection, we present results for the light-cone PDF $h_{L}(x)$ and the quasi-PDF $h_{L, \rm Q} (x)$.
\subsubsection{Light-cone PDF}
The contribution from the diagram of Fig.~(1a) can be obtained by making the replacement of $1 \rightarrow i\sigma^{+-} \gamma_{5}$ in Eq.~(\ref{e:diagram_1a_LC}). The resulting expressions with the three IR regulators are shown below.
{\flushleft For $m_{g} \neq 0$}:
\begin{eqnarray}
h^{\rm{(1a)}}_{L} (x) \Big |_{m_{g}}  &=& h^{\rm{(1a)}}_{L \rm{(s)}} (x) + h^{\rm{(1a)}}_{L \rm{(c)}} (x) \Big |_{m_{g}} \, ,
\end{eqnarray}
where the singular part of the light-cone PDF $h_L(x)$ is
\begin{eqnarray}
h^{\rm{(1a)}}_{L \rm{(s)}} (x) =
\begin{dcases}
& h^{(1a)}_{L \rm{(s)}} (x) \Big |_{m_{q}} = - \dfrac{\alpha_{s} C_{F}}{2\pi} \, \delta (x) \bigg ( {\cal P}_{\UV} + \ln \dfrac{\mu^{2}_{\UV}}{m^{2}_{q}}-1 \bigg ) \, ,
\\[0.2cm]
& h^{\rm{(1a)}}_{L \rm{(s)}} (x) \Big |_{\epsilon_{\IR}} = - \dfrac{\alpha_{s} C_{F}}{2\pi} \, \delta (x) \bigg ( {\cal P}_{\UV} - {\cal P}_{\IR} + \ln \dfrac{\mu^{2}_{\UV}}{\mu^{2}_{\IR}} \bigg ) \, , 
\end{dcases}
\end{eqnarray}
and the canonical part of the light-cone PDF $h_L(x)$ is
\begin{eqnarray}
h^{\rm{(1a)}}_{L \rm{(c)}} (x) \Big |_{m_{g}} & = & 
\dfrac{\alpha_{s} C_{F}}{2\pi} \bigg ( {\cal P}_{\UV} + \ln \dfrac{\mu^{2}_{\UV}}{x m^{2}_{g}} +  \dfrac{(1-x)(1-2x)}{x} \bigg ) \, .
\end{eqnarray}
{\flushleft For $m_{q} \neq 0$ and DR for the IR}:
\begin{eqnarray}
h^{\rm{(1a)}}_{L} (x) \Big |_{m_{q}} &=& h^{\rm{(1a)}}_{L \rm{(s)}} (x) \Big |_{m_{q}} + h^{\rm{(1a)}}_{L \rm{(c)}} (x) \Big |_{m_{q}} \nonumber \\[0.2cm]
&=& - \dfrac{\alpha_{s} C_{F}}{2\pi} \, \delta (x) \bigg ( {\cal P}_{\UV} + \ln \dfrac{\mu^{2}_{\UV}}{m^{2}_{q}}-1 \bigg ) \nonumber + \dfrac{\alpha_{s} C_{F}}{2\pi} \bigg ( {\cal P}_{\UV} + \ln \dfrac{\mu^{2}_{\UV}}{(1-x)^{2}m^{2}_{q}} + 2x -3 - \dfrac{1+x}{1-x} \bigg ) \,,
\\[0.2cm]
h^{\rm{(1a)}}_{L} (x) \Big |_{\epsilon_{\IR}}  &=& h^{\rm{(1a)}}_{L \rm{(s)}} (x) \Big |_{\epsilon_{\IR}} + h^{\rm{(1a)}}_{L \rm{(c)}} (x) \Big |_{\epsilon_{\IR}} \nonumber \\[0.2cm]
&=& - \dfrac{\alpha_{s} C_{F}}{2\pi} \, \delta (x) \bigg ( {\cal P}_{\UV} - {\cal P}_{\IR}  + \ln \dfrac{\mu^{2}_{\UV}}{\mu^{2}_{\IR}} \bigg ) + \dfrac{\alpha_{s} C_{F}}{2\pi} \, \bigg ( {\cal P}_{\UV} - {\cal P}_{\IR}  + \ln \dfrac{\mu^{2}_{\UV}}{\mu^{2}_{\IR}}  \bigg ) \, .
\end{eqnarray}
The discussions for the diagram in Fig.~(1a) made in the context of $e(x)$ carries over to $h_L(x)$. Note that the singular terms for $h_{L}(x)$ ($h_{L, \rm Q}$) and $e(x)$ ($e_{\rm Q}$) are the same except for an overall sign.  After making the replacement of $1 \rightarrow i\sigma^{+-} \gamma_{5}$ in Eq.~(\ref{e:diagram_1b_LC}), we find that the results for the diagrams shown in Figs.~(1b) and (1c) are the same as that of $e(x)$. This is due to the relevant trace algebra. As a consequence, the results of all the virtual diagrams for $h_L(x)$ are the same as that of $e(x)$. 

\subsubsection{Quasi-PDF}
The results for $h_{L, \rm{Q}}$ with the three IR regulators are shown below. 
{\flushleft{For $m_{g} \neq 0$ we have}}
\begin{eqnarray}
h^{\rm{(1a)}}_{L, \rm{Q}} (x) \Big |_{m_{g}}  &=& h^{\rm{(1a)}}_{L, \rm{Q(s)}} (x) + h^{\rm{(1a)}}_{L, \rm{Q(c)}} (x) \Big |_{m_{g}}  \, ,
\end{eqnarray}
where, for the singular part of $h_{L, \rm{Q}}$, we find
\begin{eqnarray}
h^{\rm{(1a)}}_{L, \rm{Q(s)}}  (x) =
\begin{dcases}
& h^{\rm{(1a)}}_{L, \rm{Q(s)}}  (x) \Big |_{m_{q}}  =  \dfrac{\alpha_{s} C_{F}}{2\pi}
\begin{dcases}
- \dfrac{1}{x}
& \quad x > 1  \\[0.2cm]
- \delta(x) \ln \dfrac{4p^{2}_{3}}{m^{2}_{q}} \, - \, {\rm{R}}_{0}(|x|) 
& \quad -1 < x < 1 \\[0.2cm]
\phantom{+} \dfrac{1}{x}
& \quad x < -1 \, , 
\end{dcases}
\\[0.3cm]
& h^{\rm{(1a)}}_{L, \rm{Q(s)}}  (x) \Big |_{\epsilon_{\IR}}  =  \dfrac{\alpha_{s} C_{F}}{2\pi}
\begin{dcases}
- \dfrac{1}{x}
& \quad x > 1  \\[0.2cm]
\phantom{+} \delta(x) \bigg ( {\cal P}_{\IR} - 1 - \ln \dfrac{4p^{2}_{3}}{\mu^{2}_{\IR}} \bigg ) \, - \, {\rm{R}}_{0}(|x|) 
& \quad -1 < x < 1 \\[0.2cm]
\phantom{+} \dfrac{1}{x}
& \quad x < -1 \, ,
\end{dcases}  
\end{dcases}
\end{eqnarray}
and for the canonical part of $h_{L, \rm{Q}}$ we find
\begin{eqnarray}
h_{L, \rm{Q(c)}}^{\rm{(1a)}} (x) \Big |_{m_{g}} & = &   \frac{\alpha_s C_F}{2\pi}
\begin{dcases}
\ln \frac{x}{x-1} 
& \quad  x > 1  \\[0.2cm]
\ln \dfrac{4(1-x)p^{2}_{3}}{m^{2}_{g}} + \dfrac{1-x}{x}
& \quad  0 < x < 1 \\[0.2cm]
\ln \frac{x - 1}{x}
& \quad  x < 0 \, .
\end{dcases}
\end{eqnarray}
For $m_{q} \neq 0$ and DR, we obtain
\begin{eqnarray}
h_{L, \rm{Q}}^{\rm{(1a)}} (x) \Big |_{m_{q}} & = & h_{L, \rm{Q(s)}}^{\rm{(1a)}} (x) \Big |_{m_{q}} + h_{L, \rm{Q(c)}}^{\rm{(1a)}} (x) \Big |_{m_{q}} \nonumber \\[0.2cm] 
& = & \frac{\alpha_s C_F}{2\pi}
\begin{dcases}
- \dfrac{1}{x}
&  x > 1  \\[0.2cm]
- \delta(x) \ln \dfrac{4p^{2}_{3}}{m^{2}_{q}} \, - \, {\rm{R}}_{0}(|x|) 
& -1 < x < 1 \\[0.2cm]
\phantom{+} \dfrac{1}{x}
&  x < -1 
\end{dcases}
\, + \,
\begin{dcases}
\ln \frac{x}{x-1} 
&   x > 1  \\[0.2cm]
\ln \dfrac{4x p^{2}_{3}}{(1-x)m^{2}_{q}} - \dfrac{2}{1-x}
&   0 < x < 1 \\[0.2cm]
\ln \frac{x - 1}{x}
&   x < 0 \, , 
\end{dcases}
\end{eqnarray}
\begin{eqnarray}
h_{L, \rm{Q}}^{\rm{(1a)}} (x) \Big |_{\epsilon_{\IR}} & = & h_{L, \rm{Q(s)}}^{\rm{(1a)}} (x) \Big |_{\epsilon_{\IR}} + h_{L, \rm{Q(c)}}^{\rm{(1a)}} (x) \Big |_{\epsilon_{\IR}} \nonumber \\[0.2cm]
& = & \frac{\alpha_s C_F}{2\pi}
\begin{dcases}
- \dfrac{1}{x}
&  x > 1  \\[0.2cm]
\phantom{+} \delta(x) \bigg ( {\cal P}_{\IR} - 1 - \ln \dfrac{4p^{2}_{3}}{\mu^{2}_{\IR}} \bigg ) \, - \, {\rm{R}}_{0}(|x|) 
& -1 < x < 1 \\[0.2cm]
\phantom{+} \dfrac{1}{x}
&  x < -1 
\end{dcases}  
\, + \,
\begin{dcases}
\ln \frac{x}{x-1} 
&  x > 1  \\[0.2cm]
\bigg ( - {\cal P}_{\IR} + 2(1-x) \\[0.2cm]
+ \ln \dfrac{4x(1-x)p^{2}_{3}}{\mu^{2}_{\IR}} \bigg )
&  0 < x < 1 \\[0.2cm]
\ln \frac{x - 1}{x}
&  x < 0 \, .
\end{dcases} 
\end{eqnarray}
The other diagrams yield the same results as for $e_{\rm Q}$ (see corresponding comment in previous sub-section).

\section{One-loop matching coefficient in $\msbar$} 
\label{s:problem_matching}
Schematically, the relation between light-cone and quasi-PDFs is expressed through the following factorization theorem up to power corrections that are suppressed with respect to the hadron momentum,
\begin{equation}
\tilde{q} (x; P^{3}) = \int^{+1}_{-1} \dfrac{d y}{|y|} C \bigg ( \dfrac{x}{y} \bigg ) q (y) + \mathcal{O} \bigg ( \dfrac{1}{P^{2}_{3}} \bigg ) \, .
\label{matching_def}
\end{equation}
In Eq.~(\ref{matching_def}), the symbol $\tilde{q} \, (q)$ stands for a quasi-PDF (light-cone PDF) of a parton inside a hadron, while $C$ denotes the matching coefficient. 
Like we already pointed out in the Introduction, we expect mixing with quark-gluon-quark operators, even for the quark non-singlet case. However, in this work we do not consider such mixing.
The key feature of the factorization-type formula in Eq.~(\ref{matching_def}) is the IR-finiteness of the matching coefficient $C$.
To derive the first-order correction to the matching coefficient, one applies a perturbative expansion of Eq.~(\ref{matching_def}) in powers of $\alpha_s$, leading to
\begin{equation}
C (x) = \delta(1-x) + \dfrac{\alpha_{s}C_{F}}{2\pi}  \bigg [ \widetilde{\Gamma} (x) - \Gamma (x) \bigg ] +  \dfrac{\alpha_{s}C_{F}}{2\pi} \delta (1-x)  \bigg [ \widetilde{\Pi}  - \Pi \bigg ] \, .
\label{Gamma_Pi}
\end{equation}
In Eq.~(\ref{Gamma_Pi}), $\Gamma$ ($\widetilde{\Gamma}$) and $\Pi$ ($\widetilde{\Pi}$) are the real corrections and the virtual corrections for the light-cone (quasi-) PDFs, respectively. Eq.~(\ref{Gamma_Pi}) implies that the matching coefficient, at the lowest nontrivial order in perturbation theory, is given by the difference between one-loop results for the quasi-PDFs and the light-cone PDFs.
Matching, in conjunction with proper renormalization, corrects for the different UV behavior between the light-cone and quasi-distributions such that in the limit of $P^{3} \rightarrow \infty$ one is able to recover the light-cone distributions. 

The formalism of the matching relies on the fact that the IR behavior of the light-cone and quasi distributions are the same. Previous papers on matching calculations for the twist-2 distributions have confirmed this~\cite{Xiong:2013bka,Ji:2015jwa,Ji:2015qla,Xiong:2015nua,Wang:2017qyg,Stewart:2017tvs,Radyushkin:2018cvn,Zhang:2018ggy,Izubuchi:2018srq,Liu:2018tox,Liu:2019urm,Wang:2019tgg,Wang:2019msf,Radyushkin:2019owq,Balitsky:2019krf}. 
In Ref.~\cite{Bhattacharya:2020xlt}, where we addressed the one-loop matching formula for $g_T$, we also found an agreement between the IR behavior of the light-cone PDF $g_T$ and the quasi-PDF $g_{T,\rm Q}$.
In this work, we confirm that the IR poles exactly match between $e$ ($h_L$) and $e_{\rm{Q}}$ ($h_{L, \rm{Q}}$) for all three IR regulators and for all one-loop diagrams. Finding this feature required extensive calculations. Also a non-trivial analysis was needed in the case of the singular term as discussed above in detail.

In the following, we take a brief look at the difference between one-loop results for the quasi-PDFs and the light-cone PDFs. 
This procedure gives the matching coefficient. 
For the purpose of this discussion, we take the $\MSb$ renormalized expressions of the light-cone results. As for quasi-PDFs, we renormalize the virtual diagram results in the same scheme, leaving the real diagram results as it is. The basics steps to do this exercise have been outlined in Ref.~\cite{Bhattacharya:2020xlt}.
The difference between one-loop results for $e_{\rm{Q}}(x)$ ($h_{L, \rm{Q}}$) and $e(x)$ ($h_L$) in the $\overline{\rm MS}$ scheme can be represented in the compact form
\begin{eqnarray}
C_{\overline{\mathrm{MS}}} \bigg (\xi, \dfrac{\mu^{2}}{p^{2}_{3}} \bigg ) & = & \delta (1-\xi) + C^{(\rm{s})}_{\overline{\mathrm{MS}}} \bigg (\xi, \dfrac{\mu^{2}}{p^{2}_{3}} \bigg ) + C^{(\rm{c})}_{\overline{\mathrm{MS}}} \bigg (\xi, \dfrac{\mu^{2}}{p^{2}_{3}} \bigg ) \, ,
\end{eqnarray}
where the first term corresponds to the tree-level distributions, while the second and the third terms are the differences from the singular and canonical parts of the distributions, respectively.

The difference between $e_{\rm Q}(x)$ and $e(x)$ from the singular terms reads
\begin{eqnarray}
C^{(\rm{s})}_{\overline{\mathrm{MS}}} \bigg (\xi, \dfrac{\mu^{2}}{p^{2}_{3}} \bigg ) = \dfrac{\alpha_{s} C_{F}}{2\pi}
\begin{dcases}
\phantom{+} \dfrac{1}{\xi}
& \quad \xi > 1  \\[0.2cm]
\phantom{+} \delta(\xi) \bigg ( \ln \dfrac{4p^{2}_{3}}{\mu^{2}} + 1 \bigg ) \, + \, {\rm{R}}_{0}(|\xi|) 
& \quad -1 < \xi < 1 \\[0.2cm]
- \dfrac{1}{\xi}
& \quad \xi < -1 \, .
\end{dcases}
\label{e:matching_e_sing}
\end{eqnarray}
The above equation reaffirms that there is an exact agreement in the IR poles for the singular terms between the two distributions. Moreover, both $m_{q} \neq 0$ and DR for the IR provide the very same matching coefficient. From a practical point of view, this is a very important outcome. In Eq.~(\ref{e:matching_e_sing}), we have done a change of variable $x \rightarrow \xi$ in order to reserve $x$ as the variable signifying the momentum fraction carried by quarks inside the hadrons, that is $p^{3} = x P^{3}$.
The difference between $e_{\rm Q}(x)$ and $e(x)$ from the canonical terms is
\begin{eqnarray}
C^{(\rm{c})}_{\overline{\mathrm{MS}}} \bigg (\xi, \dfrac{\mu^{2}}{p^{2}_{3}} \bigg ) & = & \dfrac{\alpha_{s} C_{F}}{2\pi}
\begin{dcases}
\bigg [ \dfrac{2}{1-\xi} \ln \dfrac{\xi}{\xi - 1} + \dfrac{1}{1-\xi} + \dfrac{1}{\xi} \bigg ]_{+} - \dfrac{1}{\xi}
& \quad \xi > 1  \\[0.2cm]
\bigg [ \dfrac{2}{1-\xi} \ln \dfrac{4\xi (1-\xi) p^{2}_{3}}{\mu^{2}} - \dfrac{1}{1-\xi} \bigg ]_{+} 
& \quad 0 < \xi < 1 \\[0.2cm]
\bigg [ \dfrac{2}{1-\xi} \ln \dfrac{\xi -1}{\xi} - \dfrac{1}{1-\xi} + \dfrac{1}{1-\xi} \bigg ]_{+} - \dfrac{1}{1- \xi}
& \quad \xi < 0 
\end{dcases}
\nonumber \\[0.2cm]
& + &  \dfrac{\alpha_{s}C_{F}}{2\pi} \delta (1-\xi) \, \bigg ( \ln \dfrac{\mu^{2}}{4p^{2}_{3}} \bigg ) \, ,
\end{eqnarray}
where the plus-prescription $[...]_{+}$ has been defined at $\xi=1$. The above equation reiterates that there is an exact agreement in the IR poles for the canonical terms of the distributions. Furthermore, just as in the case of the singular terms, this result is the same for all three IR regulators.

We now turn our attention to the difference between the one-loop results for $h_{L, \rm{Q}}(x)$ and $h_L(x)$ in the $\overline{\rm MS}$.
For the singular term we obtain a unique result,
\begin{eqnarray}
C^{(\rm{s})}_{\overline{\mathrm{MS}}} \bigg (\xi, \dfrac{\mu^{2}}{p^{2}_{3}} \bigg ) = \dfrac{\alpha_{s} C_{F}}{2\pi}
\begin{dcases}
- \dfrac{1}{\xi}
& \quad \xi > 1  \\[0.2cm]
- \delta(\xi) \bigg ( \ln \dfrac{4p^{2}_{3}}{\mu^{2}} + 1 \bigg ) \, - \, {\rm{R}}_{0}(|\xi|) 
& \quad -1 < \xi < 1 \\[0.2cm]
\phantom{+} \dfrac{1}{\xi}
& \quad \xi < -1 \, ,
\end{dcases}
\end{eqnarray}
and for the canonical term, which is also independent of the IR regulator, we get
\begin{eqnarray}
C^{(\rm{c})}_{\overline{\mathrm{MS}}} \bigg (\xi, \dfrac{\mu^{2}}{p^{2}_{3}} \bigg ) & = &  \dfrac{\alpha_{s} C_{F}}{2\pi}
\begin{dcases}
\bigg [ \dfrac{2}{1-\xi} \ln \dfrac{\xi}{\xi - 1} + \dfrac{1}{1-\xi} + \dfrac{1}{\xi} \bigg ]_{+} - \dfrac{1}{\xi}
& \quad \xi > 1  \\[0.2cm]
\bigg [ \dfrac{2}{1-\xi} \ln \dfrac{4\xi (1-\xi) p^{2}_{3}}{\mu^{2}} + 2(1-\xi) - \dfrac{1}{1-\xi} \bigg ]_{+} 
& \quad 0 < \xi < 1 \\[0.2cm]
\bigg [ \dfrac{2}{1-\xi} \ln \dfrac{\xi -1}{\xi} - \dfrac{1}{1-\xi} + \dfrac{1}{1-\xi} \bigg ]_{+} - \dfrac{1}{1- \xi}
& \quad \xi < 0 
\end{dcases}
\nonumber \\[0.2cm]
& + &  \dfrac{\alpha_{s}C_{F}}{2\pi} \delta (1-\xi) \, \bigg ( 1 + \ln \dfrac{\mu^{2}}{4p^{2}_{3}} \bigg ) \, .
\end{eqnarray}
We believe that the above results are useful, but we also repeat that operator mixing should be taken into account. As we elaborated in Ref.~\cite{Bhattacharya:2020xlt}, the problem of working with $\msbar$ matching coefficients is that the convolution integrals involved in the matching formula are divergent. The divergences can be successfully removed through an extra subtraction, for instance in the $\mmsbar$ scheme~\cite{Alexandrou:2019lfo, Bhattacharya:2020xlt}.

\section{Summary}
\label{s:conclusions}
In this paper, we present a calculation of the twist-3 light-cone PDFs $e(x)$ and $h_{L}(x)$ and their quasi-PDF counterparts $e_{\rm Q}(x)$ and $h_{L, \rm Q}(x)$ for a quark target to one-loop order in perturbation theory. 
We have regulated the IR divergences in 3 different ways: non-zero parton mass regulations, that is $m_g \neq 0$ and $m_q \neq 0$, and DR. The UV divergences are regulated using DR.

Throughout our work, we point out the main differences between these results and the ones from our previous work on $g_{T}(x)$~\cite{Bhattacharya:2020xlt}.
Specifically, we discuss the role played by singular zero-mode contributions in the matching for $e(x)$ and $h_{L}(x)$. 
While a $\delta (x)$ may or may not arise in $g_{T}(x)$ depending upon the IR scheme, it is bound to be present in $e(x)$ and $h_L(x)$. 
Even more importantly, the $\delta (x)$ in $e(x)$ and $h_L(x)$ is accompanied by prefactors that exhibit an IR divergence.
The quasi-PDFs $e_{\rm Q}(x)$ and $h_{L, \rm Q}(x)$ have a seemingly different-looking IR-pole structure at $x = 0$. However, we have shown that it is possible, in the sense of a distribution, to cast the singular terms for the quasi-PDFs into a $\delta (x)$ term whose prefactors exactly agree with those from the light-cone PDFs. This is a non-trivial point and we have provided a formal proof of this for the IR regulators $m_{q} \neq 0$ and DR. Moreover, we find that diagram-by-diagram there is an exact agreement in the IR poles between $e(x)$ and $e_{\rm Q}(x)$ as well as $h_{L}(x)$ and $h_{L, \rm Q}(x)$. This leads to the important conclusion that matching is possible for $e(x)$ and $h_{L}(x)$. Explicit results for the matching coefficients have been extracted in the $\rm{\overline{MS}}$ scheme. We repeat that complete matching equations for twist-3 PDFs most likely involve operator mixing, which we did not consider in the present work. However, we believe that our results are very useful and provide an important step toward explicitly establishing the quasi-PDF approach beyond leading twist.

\begin{acknowledgments}
We are very grateful to Yong Zhao for an important discussion about the (non-trivial) point $x = 0$ for the quasi-PDFs in DR.
The work of S.B.~and A.M.~has been supported by the National Science Foundation under grant number PHY-1812359.~A.M.~has also been supported by the U.S.~Department of Energy, Office of Science, Office of Nuclear Physics, within the framework of the TMD Topical Collaboration.~M.C. acknowledges financial support by the U.S. National Science Foundation under Grant No.\ PHY-1714407.
K.C. and A.S.\ are supported by the National Science Centre (Poland) grant SONATA BIS no.\ 2016/22/E/ST2/00013. F.S.~was funded by DFG project number 392578569.
\end{acknowledgments}



\begin{thebibliography}{99}

\bibitem{Jaffe:1991kp}
R.~Jaffe and X.~D.~Ji,
Phys. Rev. Lett. \textbf{67}, 552-555 (1991).

\bibitem{Jaffe:1991ra}
R.~Jaffe and X.~D.~Ji,
Nucl. Phys. B \textbf{375}, 527-560 (1992).

\bibitem{Balitsky:1987bk}
I.~Balitsky and V.~M.~Braun,
Nucl. Phys. B \textbf{311}, 541-584 (1989).

\bibitem{Kanazawa:2015ajw}
K.~Kanazawa, Y.~Koike, A.~Metz, D.~Pitonyak and M.~Schlegel,
Phys. Rev. D \textbf{93}, 054024 (2016)
[arXiv:1512.07233 [hep-ph]].

\bibitem{Burkardt:2008ps}
M.~Burkardt,
Phys. Rev. D \textbf{88}, 114502 (2013)
[arXiv:0810.3589 [hep-ph]].

\bibitem{Seng:2018wwp}
C.~Y.~Seng,
Phys. Rev. Lett. \textbf{122}, 072001 (2019)
[arXiv:1809.00307 [hep-ph]].

\bibitem{Hatta:2020iin}
Y.~Hatta and Y.~Zhao,
[arXiv:2006.02798 [hep-ph]].

\bibitem{Ji:2020baz}
X.~Ji,
[arXiv:2003.04478 [hep-ph]].

\bibitem{Levelt:1994np}
J.~Levelt and P.~Mulders,
Phys. Lett. B \textbf{338}, 357-362 (1994)
[arXiv:hep-ph/9408257 [hep-ph]].

\bibitem{Airapetian:1999tv}
A.~Airapetian \textit{et al.} [HERMES],
Phys. Rev. Lett. \textbf{84}, 4047-4051 (2000)
[arXiv:hep-ex/9910062 [hep-ex]].

\bibitem{Airapetian:2001eg}
A.~Airapetian \textit{et al.} [HERMES],
Phys. Rev. D \textbf{64}, 097101 (2001)
[arXiv:hep-ex/0104005 [hep-ex]].

\bibitem{Avakian:2003pk}
H.~Avakian \textit{et al.} [CLAS],
Phys. Rev. D \textbf{69}, 112004 (2004)
[arXiv:hep-ex/0301005 [hep-ex]].

\bibitem{Gohn:2014zbz}
W.~Gohn \textit{et al.} [CLAS],
Phys. Rev. D \textbf{89}, 072011 (2014)
[arXiv:1402.4097 [hep-ex]].

\bibitem{Bacchetta:2003vn}
A.~Bacchetta and M.~Radici,
Phys. Rev. D \textbf{69}, 074026 (2004)
[arXiv:hep-ph/0311173 [hep-ph]].

\bibitem{Courtoy:2014ixa}
A.~Courtoy,
[arXiv:1405.7659 [hep-ph]].

\bibitem{Tangerman:1994bb}
R.~Tangerman and P.~Mulders,
[arXiv:hep-ph/9408305 [hep-ph]].

\bibitem{Koike:2008du}
Y.~Koike, K.~Tanaka and S.~Yoshida,
Phys. Lett. B \textbf{668}, 286-292 (2008)
[arXiv:0805.2289 [hep-ph]].

\bibitem{Liang:2012rb}
Z.~T.~Liang, A.~Metz, D.~Pitonyak, A.~Sch{\"a}fer, Y.~K.~Song and J.~Zhou,
Phys. Lett. B \textbf{712}, 235-239 (2012)
[arXiv:1203.3956 [hep-ph]].

\bibitem{Metz:2012fq}
A.~Metz, D.~Pitonyak, A.~Sch{\"a}fer and J.~Zhou,
Phys. Rev. D \textbf{86}, 114020 (2012)
[arXiv:1210.6555 [hep-ph]].

\bibitem{Burkardt:1995ts}
M.~Burkardt,
Phys. Rev. D \textbf{52}, 3841-3852 (1995)
[arXiv:hep-ph/9505226 [hep-ph]].

\bibitem{Burkhardt:1970ti}
H.~Burkhardt and W.~Cottingham,
Annals Phys. \textbf{56}, 453-463 (1970)

\bibitem{Kodaira:1998jn}
J.~Kodaira and K.~Tanaka,
Prog. Theor. Phys. \textbf{101}, 191-242 (1999)
[arXiv:hep-ph/9812449 [hep-ph]].

\bibitem{Efremov:2002qh}
A.~Efremov and P.~Schweitzer,
JHEP \textbf{08}, 006 (2003)
[arXiv:hep-ph/0212044 [hep-ph]].

\bibitem{Pasquini:2018oyz}
B.~Pasquini and S.~Rodini,
Phys. Lett. B \textbf{788}, 414-424 (2019)
doi:10.1016/j.physletb.2018.11.033
[arXiv:1806.10932 [hep-ph]].

\bibitem{Ma:2020kjz}
J.~Ma and G.~Zhang,
[arXiv:2003.13920 [hep-ph]].

\bibitem{Burkardt:2001iy}
M.~Burkardt and Y.~Koike,
Nucl. Phys. B \textbf{632}, 311-329 (2002)
[arXiv:hep-ph/0111343 [hep-ph]].

\bibitem{Signal:1996ct}
A.~Signal,
Nucl. Phys. B \textbf{497}, 415-434 (1997)
[arXiv:hep-ph/9610480 [hep-ph]].

\bibitem{Jakob:1997wg}
R.~Jakob, P.~Mulders and J.~Rodrigues,
Nucl. Phys. A \textbf{626}, 937-965 (1997)
[arXiv:hep-ph/9704335 [hep-ph]].

\bibitem{Aslan:2018tff} 
F.~Aslan and M.~Burkardt,
Phys.\ Rev.\ D {\bf 101}, 016010 (2020)
[arXiv:1811.00938 [nucl-th]].
  
\bibitem{Wakamatsu:2003uu}
M.~Wakamatsu and Y.~Ohnishi,
Phys. Rev. D \textbf{67}, 114011 (2003)
[arXiv:hep-ph/0303007 [hep-ph]].
  
\bibitem{Ohnishi:2003mf}
Y.~Ohnishi and M.~Wakamatsu,
Phys. Rev. D \textbf{69}, 114002 (2004)
[arXiv:hep-ph/0312044 [hep-ph]].
  
\bibitem{Cebulla:2007ej}
C.~Cebulla, J.~Ossmann, P.~Schweitzer and D.~Urbano,
Acta Phys. Polon. B \textbf{39}, 609-640 (2008)
[arXiv:0710.3103 [hep-ph]].

\bibitem{Schweitzer:2003uy}
P.~Schweitzer,
Phys. Rev. D \textbf{67}, 114010 (2003)
[arXiv:hep-ph/0303011 [hep-ph]].

\bibitem{Mukherjee:2009uy}
A.~Mukherjee,
Phys. Lett. B \textbf{687}, 180-183 (2010)
[arXiv:0912.1446 [hep-ph]].

\bibitem{Kundu:2001pk}
R.~Kundu and A.~Metz,
Phys. Rev. D \textbf{65}, 014009 (2002)
[arXiv:hep-ph/0107073 [hep-ph]].

\bibitem{Aslan:2018zzk}
F.~Aslan, M.~Burkardt, C.~Lorc{\'e}, A.~Metz and B.~Pasquini,
Phys. Rev. D \textbf{98}, 014038 (2018)
[arXiv:1802.06243 [hep-ph]].

\bibitem{Ji:2013dva} 
X.~Ji,
Phys.\ Rev.\ Lett.\  {\bf 110}, 262002 (2013)
[arXiv:1305.1539 [hep-ph]].

\bibitem{Ji:2014gla} 
X.~Ji,
Sci.\ China Phys.\ Mech.\ Astron.\  {\bf 57}, 1407 (2014)
[arXiv:1404.6680 [hep-ph]].

\bibitem{Xiong:2013bka} 
X.~Xiong, X.~Ji, J.~H.~Zhang and Y.~Zhao,
Phys.\ Rev.\ D {\bf 90}, 014051 (2014)
[arXiv:1310.7471 [hep-ph]].

\bibitem{Stewart:2017tvs} 
I.~W.~Stewart and Y.~Zhao,
Phys.\ Rev.\ D {\bf 97}, 054512 (2018)
[arXiv:1709.04933 [hep-ph]].

\bibitem{Izubuchi:2018srq} 
T.~Izubuchi, X.~Ji, L.~Jin, I.~W.~Stewart and Y.~Zhao,
Phys.\ Rev.\ D {\bf 98}, 056004 (2018)
[arXiv:1801.03917 [hep-ph]].

\bibitem{Braun:1994jq} 
V.~Braun, P.~Gornicki and L.~Mankiewicz,
Phys.\ Rev.\ D {\bf 51}, 6036 (1995)
[hep-ph/9410318].

\bibitem{Detmold:2005gg} 
W.~Detmold and C.~J.~D.~Lin,
Phys.\ Rev.\ D {\bf 73}, 014501 (2006)
[hep-lat/0507007].

\bibitem{Braun:2007wv} 
V.~Braun and D.~Mueller,
Eur.\ Phys.\ J.\ C {\bf 55}, 349 (2008)
[arXiv:0709.1348 [hep-ph]].

\bibitem{Ma:2014jla} 
Y.~Q.~Ma and J.~W.~Qiu,
Phys.\ Rev.\ D {\bf 98}, 074021 (2018)
[arXiv:1404.6860 [hep-ph]].

\bibitem{Chambers:2017dov} 
A.~J.~Chambers {\it et al.},
Phys.\ Rev.\ Lett.\  {\bf 118}, 242001 (2017)
[arXiv:1703.01153 [hep-lat]].

\bibitem{Hansen:2017mnd} 
M.~T.~Hansen, H.~B.~Meyer and D.~Robaina,
Phys.\ Rev.\ D {\bf 96}, 094513 (2017)
[arXiv:1704.08993 [hep-lat]].

\bibitem{Radyushkin:2017cyf} 
A.~V.~Radyushkin,
Phys.\ Rev.\ D {\bf 96}, 034025 (2017)
[arXiv:1705.01488 [hep-ph]].
  
\bibitem{Orginos:2017kos} 
K.~Orginos, A.~Radyushkin, J.~Karpie and S.~Zafeiropoulos,
Phys.\ Rev.\ D {\bf 96}, 094503 (2017)
[arXiv:1706.05373 [hep-ph]].

\bibitem{Ma:2017pxb} 
Y.~Q.~Ma and J.~W.~Qiu,
Phys.\ Rev.\ Lett.\  {\bf 120}, 022003 (2018)
[arXiv:1709.03018 [hep-ph]].

\bibitem{Radyushkin:2017lvu} 
A.~V.~Radyushkin,
Phys.\ Lett.\ B {\bf 781}, 433 (2018)
[arXiv:1710.08813 [hep-ph]].

\bibitem{Liang:2017mye} 
J.~Liang, K.~F.~Liu and Y.~B.~Yang,
EPJ Web Conf.\  {\bf 175}, 14014 (2018)
[arXiv:1710.11145 [hep-lat]].

\bibitem{Detmold:2018kwu} 
W.~Detmold, I.~Kanamori, C.~J.~D.~Lin, S.~Mondal and Y.~Zhao,
arXiv:1810.12194 [hep-lat].
 
\bibitem{Ji:2015jwa} 
X.~Ji and J.~H.~Zhang,
Phys.\ Rev.\ D {\bf 92}, 034006 (2015)
[arXiv:1505.07699 [hep-ph]].
  
\bibitem{Ishikawa:2016znu} 
T.~Ishikawa, Y.~Q.~Ma, J.~W.~Qiu and S.~Yoshida,
arXiv:1609.02018 [hep-lat].

\bibitem{Chen:2016fxx} 
J.~W.~Chen, X.~Ji and J.~H.~Zhang,
Nucl.\ Phys.\ B {\bf 915}, 1 (2017)
[arXiv:1609.08102 [hep-ph]].
  
\bibitem{Constantinou:2017sej} 
M.~Constantinou and H.~Panagopoulos,
Phys.\ Rev.\ D {\bf 96}, 054506 (2017)
[arXiv:1705.11193 [hep-lat]].

\bibitem{Alexandrou:2017huk} 
C.~Alexandrou, K.~Cichy, M.~Constantinou, K.~Hadjiyiannakou, K.~Jansen, H.~Panagopoulos and F.~Steffens,
Nucl.\ Phys.\ B {\bf 923}, 394 (2017)
[arXiv:1706.00265 [hep-lat]].

\bibitem{Chen:2017mzz} 
J.~W.~Chen, T.~Ishikawa, L.~Jin, H.~W.~Lin, Y.~B.~Yang, J.~H.~Zhang and Y.~Zhao,
Phys.\ Rev.\ D {\bf 97}, 014505 (2018)
[arXiv:1706.01295 [hep-lat]].

\bibitem{Ji:2017oey} 
X.~Ji, J.~H.~Zhang and Y.~Zhao,
Phys.\ Rev.\ Lett.\  {\bf 120}, 112001 (2018)
[arXiv:1706.08962 [hep-ph]].

\bibitem{Ishikawa:2017faj} 
T.~Ishikawa, Y.~Q.~Ma, J.~W.~Qiu and S.~Yoshida,
Phys.\ Rev.\ D {\bf 96}, 094019 (2017)
[arXiv:1707.03107 [hep-ph]].

\bibitem{Green:2017xeu} 
J.~Green, K.~Jansen and F.~Steffens,
Phys.\ Rev.\ Lett.\  {\bf 121}, 022004 (2018)
[arXiv:1707.07152 [hep-lat]].


\bibitem{Spanoudes:2018zya} 
G.~Spanoudes and H.~Panagopoulos,
Phys.\ Rev.\ D {\bf 98}, 014509 (2018)
[arXiv:1805.01164 [hep-lat]].
  
\bibitem{Zhang:2018diq} 
J.~H.~Zhang, X.~Ji, A.~Sch\"afer, W.~Wang and S.~Zhao,
arXiv:1808.10824 [hep-ph].

\bibitem{Li:2018tpe} 
Z.~Y.~Li, Y.~Q.~Ma and J.~W.~Qiu,
Phys.\ Rev.\ Lett.\  {\bf 122}, 062002 (2019)
[arXiv:1809.01836 [hep-ph]].

\bibitem{Constantinou:2019vyb} 
M.~Constantinou, H.~Panagopoulos and G.~Spanoudes,
arXiv:1901.03862 [hep-lat].
  
\bibitem{Wang:2017qyg}
W.~Wang, S.~Zhao and R.~Zhu,
Eur. Phys. J. C \textbf{78}, 147 (2018)
[arXiv:1708.02458 [hep-ph]].

\bibitem{Ji:2014hxa} 
X.~Ji, P.~Sun, X.~Xiong and F.~Yuan,
Phys.\ Rev.\ D {\bf 91}, 074009 (2015)
[arXiv:1405.7640 [hep-ph]].

\bibitem{Li:2016amo} 
H.~n.~Li,
Phys.\ Rev.\ D {\bf 94}, 074036 (2016)
[arXiv:1602.07575 [hep-ph]].
  
\bibitem{Monahan:2016bvm} 
C.~Monahan and K.~Orginos,
JHEP {\bf 1703}, 116 (2017)
[arXiv:1612.01584 [hep-lat]].
  
\bibitem{Radyushkin:2016hsy} 
A.~Radyushkin,
Phys.\ Lett.\ B {\bf 767}, 314 (2017)
[arXiv:1612.05170 [hep-ph]].
  
\bibitem{Radyushkin:2017ffo} 
A.~Radyushkin,
Phys.\ Lett.\ B {\bf 770}, 514 (2017)
[arXiv:1702.01726 [hep-ph]].
  
\bibitem{Carlson:2017gpk} 
C.~E.~Carlson and M.~Freid,
Phys.\ Rev.\ D {\bf 95}, 094504 (2017)
[arXiv:1702.05775 [hep-ph]].
  
\bibitem{Briceno:2017cpo} 
R.~A.~Brice\~no, M.~T.~Hansen and C.~J.~Monahan,
Phys.\ Rev.\ D {\bf 96}, 014502 (2017)
[arXiv:1703.06072 [hep-lat]].
  
\bibitem{Xiong:2017jtn} 
X.~Xiong, T.~Luu and U.~G.~Mei{\ss}ner,
arXiv:1705.00246 [hep-ph].
  
\bibitem{Rossi:2017muf} 
G.~C.~Rossi and M.~Testa,
Phys.\ Rev.\ D {\bf 96}, 014507 (2017)
[arXiv:1706.04428 [hep-lat]].
  
\bibitem{Ji:2017rah} 
X.~Ji, J.~H.~Zhang and Y.~Zhao,
Nucl.\ Phys.\ B {\bf 924}, 366 (2017)
[arXiv:1706.07416 [hep-ph]].
  
\bibitem{Chen:2017mie} 
J.~W.~Chen, T.~Ishikawa, L.~Jin, H.~W.~Lin, Y.~B.~Yang, J.~H.~Zhang and Y.~Zhao,
arXiv:1710.01089 [hep-lat].
  
\bibitem{Monahan:2017hpu} 
C.~Monahan,
Phys.\ Rev.\ D {\bf 97}, 054507 (2018)
[arXiv:1710.04607 [hep-lat]].

\bibitem{Radyushkin:2018cvn} 
A.~Radyushkin,
Phys.\ Rev.\ D {\bf 98}, 014019 (2018)
[arXiv:1801.02427 [hep-ph]].

\bibitem{Zhang:2018ggy} 
J.~H.~Zhang, J.~W.~Chen and C.~Monahan,
Phys.\ Rev.\ D {\bf 97}, 074508 (2018)
[arXiv:1801.03023 [hep-ph]].

\bibitem{Ji:2018hvs} 
X.~Ji, L.~C.~Jin, F.~Yuan, J.~H.~Zhang and Y.~Zhao,
arXiv:1801.05930 [hep-ph].

\bibitem{Xu:2018mpf} 
J.~Xu, Q.~A.~Zhang and S.~Zhao,
Phys.\ Rev.\ D {\bf 97}, 114026 (2018)
[arXiv:1804.01042 [hep-ph]].

\bibitem{Jia:2018qee} 
Y.~Jia, S.~Liang, X.~Xiong and R.~Yu,
Phys.\ Rev.\ D {\bf 98}, 054011 (2018)
[arXiv:1804.04644 [hep-th]].

\bibitem{Briceno:2018lfj} 
R.~A.~Brice\~no, J.~V.~Guerrero, M.~T.~Hansen and C.~J.~Monahan,
Phys.\ Rev.\ D {\bf 98}, 014511 (2018)
[arXiv:1805.01034 [hep-lat]].

\bibitem{Rossi:2018zkn} 
G.~Rossi and M.~Testa,
Phys.\ Rev.\ D {\bf 98}, 054028 (2018)
[arXiv:1806.00808 [hep-lat]].

\bibitem{Radyushkin:2018nbf} 
A.~V.~Radyushkin,
Phys.\ Lett.\ B {\bf 788}, 380 (2019)
[arXiv:1807.07509 [hep-ph]].

\bibitem{Ji:2018waw} 
X.~Ji, Y.~Liu and I.~Zahed,
arXiv:1807.07528 [hep-ph].

\bibitem{Karpie:2018zaz} 
J.~Karpie, K.~Orginos and S.~Zafeiropoulos,
JHEP {\bf 1811}, 178 (2018)
[arXiv:1807.10933 [hep-lat]].

\bibitem{Braun:2018brg} 
V.~M.~Braun, A.~Vladimirov and J.~H.~Zhang,
Phys.\ Rev.\ D {\bf 99}, 014013 (2019)
[arXiv:1810.00048 [hep-ph]].

\bibitem{Liu:2018tox} 
Y.~S.~Liu, W.~Wang, J.~Xu, Q.~A.~Zhang, S.~Zhao and Y.~Zhao,
arXiv:1810.10879 [hep-ph].

\bibitem{Ebert:2018gzl} 
M.~A.~Ebert, I.~W.~Stewart and Y.~Zhao,
Phys.\ Rev.\ D {\bf 99}, 034505 (2019)
[arXiv:1811.00026 [hep-ph]].
  
\bibitem{Briceno:2018qfa} 
R.~A.~Brice\~no, J.~V.~Guerrero, M.~T.~Hansen and C.~J.~Monahan,
PoS LATTICE {\bf 2018}, 111 (2018)
[arXiv:1811.01286 [hep-lat]].

\bibitem{Ebert:2019okf}
M.~A.~Ebert, I.~W.~Stewart and Y.~Zhao,
JHEP \textbf{09} (2019), 037
doi:10.1007/JHEP09(2019)037
[arXiv:1901.03685 [hep-ph]].
  
\bibitem{Lin:2014zya} 
H.~W.~Lin, J.~W.~Chen, S.~D.~Cohen and X.~Ji,
Phys.\ Rev.\ D {\bf 91}, 054510 (2015)
[arXiv:1402.1462 [hep-ph]].

\bibitem{Alexandrou:2015rja} 
C.~Alexandrou, K.~Cichy, V.~Drach, E.~Garcia-Ramos, K.~Hadjiyiannakou, K.~Jansen, F.~Steffens and C.~Wiese,
Phys.\ Rev.\ D {\bf 92}, 014502 (2015)
[arXiv:1504.07455 [hep-lat]].

\bibitem{Chen:2016utp} 
J.~W.~Chen, S.~D.~Cohen, X.~Ji, H.~W.~Lin and J.~H.~Zhang,
Nucl.\ Phys.\ B {\bf 911}, 246 (2016)
[arXiv:1603.06664 [hep-ph]].

\bibitem{Alexandrou:2016jqi} 
C.~Alexandrou, K.~Cichy, M.~Constantinou, K.~Hadjiyiannakou, K.~Jansen, F.~Steffens and C.~Wiese,
Phys.\ Rev.\ D {\bf 96}, 014513 (2017)
[arXiv:1610.03689 [hep-lat]].

\bibitem{Zhang:2017bzy} 
J.~H.~Zhang, J.~W.~Chen, X.~Ji, L.~Jin and H.~W.~Lin,
Phys.\ Rev.\ D {\bf 95}, 094514 (2017)
[arXiv:1702.00008 [hep-lat]].

\bibitem{Lin:2017ani} 
H.~W.~Lin {\it et al.} [LP3 Collaboration],
Phys.\ Rev.\ D {\bf 98}, 054504 (2018)
[arXiv:1708.05301 [hep-lat]].

\bibitem{Bali:2017gfr} 
G.~S.~Bali {\it et al.},
Eur.\ Phys.\ J.\ C {\bf 78}, 217 (2018)
[arXiv:1709.04325 [hep-lat]].

\bibitem{Alexandrou:2017dzj} 
C.~Alexandrou {\it et al.},
EPJ Web Conf.\  {\bf 175}, 14008 (2018)
[arXiv:1710.06408 [hep-lat]].

\bibitem{Chen:2017gck} 
J.~H.~Zhang {\it et al.} [LP3 Collaboration],
Nucl.\ Phys.\ B {\bf 939}, 429 (2019)
[arXiv:1712.10025 [hep-ph]].

\bibitem{Alexandrou:2018pbm} 
C.~Alexandrou, K.~Cichy, M.~Constantinou, K.~Jansen, A.~Scapellato and F.~Steffens,
Phys.\ Rev.\ Lett.\  {\bf 121}, 112001 (2018)
[arXiv:1803.02685 [hep-lat]].

\bibitem{Chen:2018fwa} 
J.~W.~Chen, L.~Jin, H.~W.~Lin, Y.~S.~Liu, A.~Sch{\"a}fer, Y.~B.~Yang, J.~H.~Zhang and Y.~Zhao,
Phys.\ Rev.\ D \textbf{100}, 034505 (2019)
arXiv:1804.01483 [hep-lat].

\bibitem{Alexandrou:2018eet} 
C.~Alexandrou, K.~Cichy, M.~Constantinou, K.~Jansen, A.~Scapellato and F.~Steffens,
Phys.\ Rev.\ D {\bf 98}, 091503 (2018)
[arXiv:1807.00232 [hep-lat]].

\bibitem{Liu:2018uuj} 
Y.~S.~Liu, J.~W.~Chen, L.~Jin, H.~W.~Lin, Y.~B.~Yang, J.~H.~Zhang and Y.~Zhao,
Phys.\ Rev.\ D \textbf{101}, 034020 (2020)
arXiv:1807.06566 [hep-lat].

\bibitem{Bali:2018spj} 
G.~S.~Bali {\it et al.},
Phys.\ Rev.\ D {\bf 98}, 094507 (2018)
[arXiv:1807.06671 [hep-lat]].

\bibitem{Lin:2018qky} 
H.~W.~Lin {\it et al.},
Phys.\ Rev.\ Lett.\  {\bf 121}, 242003 (2018)
[arXiv:1807.07431 [hep-lat]].

\bibitem{Fan:2018dxu} 
Z.~Y.~Fan, Y.~B.~Yang, A.~Anthony, H.~W.~Lin and K.~F.~Liu,
Phys.\ Rev.\ Lett.\  {\bf 121}, 242001 (2018)
[arXiv:1808.02077 [hep-lat]].

\bibitem{Bali:2018qat} 
G.~S.~Bali, V.~M.~Braun, M.~G\"ockeler, M.~Gruber, F.~Hutzler, P.~Korcyl, A.~Sch\"afer and P.~Wein,
PoS \textbf{LATTICE2018} 107 (2018)
arXiv:1811.06050 [hep-lat].

\bibitem{Sufian:2019bol} 
R.~S.~Sufian, J.~Karpie, C.~Egerer, K.~Orginos, J.~W.~Qiu and D.~G.~Richards,
Phys.\ Rev.\ D \textbf{99}, 074507 (2019)
arXiv:1901.03921 [hep-lat].

\bibitem{Alexandrou:2019lfo} 
C.~Alexandrou, K.~Cichy, M.~Constantinou, K.~Hadjiyiannakou, K.~Jansen, A.~Scapellato and F.~Steffens,
Phys.\ Rev.\ D {\bf 99}, 114504 (2019)
arXiv:1902.00587 [hep-lat].

\bibitem{Izubuchi:2019lyk}
T.~Izubuchi, L.~Jin, C.~Kallidonis, N.~Karthik, S.~Mukherjee, P.~Petreczky, C.~Shugert and S.~Syritsyn,
Phys.\ Rev.\ D \textbf{100}, 034516 (2019)
[arXiv:1905.06349 [hep-lat]].

\bibitem{Cichy:2019ebf}
K.~Cichy, L.~Del Debbio and T.~Giani,
JHEP \textbf{1910}, 137 (2019)
[arXiv:1907.06037 [hep-ph]].

\bibitem{Joo:2019jct}
B.~Jo\'o, J.~Karpie, K.~Orginos, A.~Radyushkin, D.~Richards and S.~Zafeiropoulos,
JHEP \textbf{1912}, 081 (2019)
[arXiv:1908.09771 [hep-lat]].

\bibitem{Joo:2019bzr}
B.~Jo{\'o}, J.~Karpie, K.~Orginos, A.~V.~Radyushkin, D.~G.~Richards, R.~S.~Sufian and S.~Zafeiropoulos,
Phys.\ Rev.\ D \textbf{100}, 114512 (2019)
[arXiv:1909.08517 [hep-lat]].

\bibitem{Alexandrou:2019dax}
C.~Alexandrou, K.~Cichy, M.~Constantinou, K.~Hadjiyiannakou, K.~Jansen, A.~Scapellato and F.~Steffens,
[arXiv:1910.13229 [hep-lat]].

\bibitem{Chai:2020nxw}
Y.~Chai, Y.~Li, S.~Xia, C.~Alexandrou, K.~Cichy, M.~Constantinou, X.~Feng, K.~Hadjiyiannakou, K.~Jansen, G.~Koutsou, C.~Liu, A.~Scapellato and F.~Steffens,
[arXiv:2002.12044 [hep-lat]].

\bibitem{Joo:2020spy}
B.~Jo{\'o}, J.~Karpie, K.~Orginos, A.~V.~Radyushkin, D.~G.~Richards and S.~Zafeiropoulos,
[arXiv:2004.01687 [hep-lat]].

\bibitem{Bhat:2020ktg}
M.~Bhat, K.~Cichy, M.~Constantinou and A.~Scapellato,
[arXiv:2005.02102 [hep-lat]].

\bibitem{Gamberg:2014zwa} 
L.~Gamberg, Z.~B.~Kang, I.~Vitev and H.~Xing,
Phys.\ Lett.\ B {\bf 743}, 112 (2015)
[arXiv:1412.3401 [hep-ph]].

\bibitem{Bacchetta:2016zjm} 
A.~Bacchetta, M.~Radici, B.~Pasquini and X.~Xiong,
Phys.\ Rev.\ D {\bf 95}, 014036 (2017)
[arXiv:1608.07638 [hep-ph]].

\bibitem{Nam:2017gzm} 
S.~i.~Nam,
Mod.\ Phys.\ Lett.\ A {\bf 32}, 1750218 (2017)
[arXiv:1704.03824 [hep-ph]].

\bibitem{Broniowski:2017wbr} 
W.~Broniowski and E.~Ruiz Arriola,
Phys.\ Lett.\ B {\bf 773}, 385 (2017)
[arXiv:1707.09588 [hep-ph]].

\bibitem{Hobbs:2017xtq} 
T.~J.~Hobbs,
Phys.\ Rev.\ D {\bf 97}, 054028 (2018)
[arXiv:1708.05463 [hep-ph]].

\bibitem{Broniowski:2017gfp} 
W.~Broniowski and E.~Ruiz Arriola,
Phys.\ Rev.\ D {\bf 97}, 034031 (2018)
[arXiv:1711.03377 [hep-ph]].

\bibitem{Xu:2018eii} 
S.~S.~Xu, L.~Chang, C.~D.~Roberts and H.~S.~Zong,
Phys.\ Rev.\ D {\bf 97}, 094014 (2018)
[arXiv:1802.09552 [nucl-th]].

\bibitem{Bhattacharya:2018zxi} 
S.~Bhattacharya, C.~Cocuzza and A.~Metz,
Phys.\ Lett.\ B {\bf 788}, 453 (2019)
[arXiv:1808.01437 [hep-ph]].
  
\bibitem{Bhattacharya:2019cme}
S.~Bhattacharya, C.~Cocuzza and A.~Metz,
Phys. Rev. D \textbf{102}, 054021 (2020)
[arXiv:1903.05721 [hep-ph]].

\bibitem{Son:2019ghf}
H.~D.~Son, A.~Tandogan and M.~V.~Polyakov,
[arXiv:1911.01955 [hep-ph]].

\bibitem{Ma:2019agv}
Z.~L.~Ma, J.~Q.~Zhu and Z.~Lu,
[arXiv:1912.12816 [hep-ph]].

\bibitem{Kock:2020frx}
A.~Kock, Y.~Liu and I.~Zahed,
[arXiv:2004.01595 [hep-ph]].

\bibitem{Luo:2020yqj}
X.~Luo and H.~Sun,
[arXiv:2005.09832 [hep-ph]].

\bibitem{Monahan:2018euv} 
C.~Monahan,
PoS LATTICE {\bf 2018}, 018 (2018)
[arXiv:1811.00678 [hep-lat]].

\bibitem{Cichy:2018mum} 
K.~Cichy and M.~Constantinou,
Adv.\ High Energy Phys.\ \textbf{2019}, 3036904 (2019)
arXiv:1811.07248 [hep-lat].

\bibitem{Zhao:2018fyu} 
Y.~Zhao,
Int.\ J.\ Mod.\ Phys.\ A {\bf 33}, 1830033 (2019)
[arXiv:1812.07192 [hep-ph]].

\bibitem{Ji:2020ect}
X.~Ji, Y.~S.~Liu, Y.~Liu, J.~H.~Zhang and Y.~Zhao,
[arXiv:2004.03543 [hep-ph]].

\bibitem{Constantinou:2020pek}
M.~Constantinou,
[arXiv:2010.02445 [hep-lat]].


\bibitem{Ji:2015qla}
X.~Ji, A.~Sch\"afer, X.~Xiong and J.~H.~Zhang,
Phys. Rev. D \textbf{92}, 014039 (2015)
[arXiv:1506.00248 [hep-ph]].

\bibitem{Xiong:2015nua}
X.~Xiong and J.~H.~Zhang,
Phys. Rev. D \textbf{92}, 054037 (2015)
[arXiv:1509.08016 [hep-ph]].

\bibitem{Liu:2019urm}
Y.~S.~Liu, W.~Wang, J.~Xu, Q.~A.~Zhang, J.~H.~Zhang, S.~Zhao and Y.~Zhao,
Phys. Rev. D \textbf{100}, 034006 (2019)
[arXiv:1902.00307 [hep-ph]].

\bibitem{Wang:2019tgg}
W.~Wang, J.~H.~Zhang, S.~Zhao and R.~Zhu,
Phys. Rev. D \textbf{100}, 074509 (2019)
[arXiv:1904.00978 [hep-ph]].

\bibitem{Wang:2019msf}
W.~Wang, Y.~M.~Wang, J.~Xu and S.~Zhao,
[arXiv:1908.09933 [hep-ph]].

\bibitem{Radyushkin:2019owq}
A.~V.~Radyushkin,
Phys. Rev. D \textbf{100}, 116011 (2019)
[arXiv:1909.08474 [hep-ph]].

\bibitem{Balitsky:2019krf}
I.~Balitsky, W.~Morris and A.~Radyushkin,
[arXiv:1910.13963 [hep-ph]].

\bibitem{Bhattacharya:2020xlt}
S.~Bhattacharya, K.~Cichy, M.~Constantinou, A.~Metz, A.~Scapellato and F.~Steffens,
Phys. Rev. D \textbf{102}, 034005 (2020)
[arXiv:2005.10939 [hep-ph]].

\bibitem{Bhattacharya:2020cen}
S.~Bhattacharya, K.~Cichy, M.~Constantinou, A.~Metz, A.~Scapellato and F.~Steffens,
[arXiv:2004.04130 [hep-lat]].

\bibitem{Yan:1973qg}
T.~M.~Yan,
Phys. Rev. D \textbf{7}, 1780-1800 (1973).

\bibitem{Ligeti:2008ac}
Z.~Ligeti, I.~W.~Stewart and F.~J.~Tackmann,
Phys. Rev. D \textbf{78}, 114014 (2008).


\end{thebibliography}
\end{document}